\documentclass{article} % For LaTeX2e
\usepackage{iclr2021_conference,times}

\usepackage{amsmath,amsfonts,bm}

\def\eqref#1{equation~\ref{#1}}
\def\Eqref#1{Equation~\ref{#1}}
\def\1{\bm{1}}

\DeclareMathAlphabet{\mathsfit}{\encodingdefault}{\sfdefault}{m}{sl}
\SetMathAlphabet{\mathsfit}{bold}{\encodingdefault}{\sfdefault}{bx}{n}

\usepackage{hyperref}
\usepackage{url}

\usepackage{xcolor}
\usepackage{colortbl}
\usepackage{multirow}
\usepackage{booktabs}
\usepackage{wrapfig}
\usepackage{graphicx}
\usepackage{subcaption}
\usepackage{amssymb}
\usepackage{bm}
\usepackage[font=small,skip=2pt]{caption}

\usepackage{mdwlist}
\usepackage{paralist}
\usepackage{threeparttable}
\usepackage{tablefootnote}
\usepackage[ruled,linesnumbered,vlined]{algorithm2e}
\usepackage[textwidth=1.3in,textsize=footnotesize]{todonotes}
\usepackage{appendix}

\definecolor{mygreen}{RGB}{20, 90, 50}
\definecolor{yyblue}{RGB}{176, 224, 230}

\newcommand{\method}{MARS\xspace}

\newcommand{\hide}[1]{} %hide the content

\title{\method: Markov Molecular Sampling for\\ Multi-objective Drug Discovery}

\author{Yutong Xie$^{\dagger\diamond}$, Chence Shi$^{\dagger\triangle}$, Hao Zhou$^\dagger$$^{*}$, Yuwei Yang$^\dagger$, Weinan Zhang$^\ddagger$, Yong Yu$^\ddagger$, Lei Li$^\dagger$\thanks{Work was done while Yutong Xie and Chence Shi were research interns at ByteDance. Corresponding to:  \href{mailto:zhouhao.nlp@bytedance.com}{zhouhao.nlp@bytedance.com}  and \href{mailto:lileilab@bytedance.com}{lileilab@bytedance.com}. } \\
$^\dagger$ByteDance AI Lab, Shanghai, China \\
$^\diamond$University of Michigan, Ann Arbor, MI, USA \\
$^\triangle$Montr\'eal Institute of Learning Algorithms, Montreal, Canada \\
$^\ddagger$Department of Computer Science and Engineering, Shanghai Jiao Tong University, China \\
}

\iclrfinalcopy % Uncomment for camera-ready version, but NOT for submission.

\begin{document}

\maketitle

\begin{abstract}
    Searching for novel molecules with desired chemical properties is crucial in drug discovery. 
Existing work focuses on developing neural models to generate either molecular sequences or chemical graphs. 
However, it remains a big challenge to find novel and diverse compounds satisfying several properties. 
In this paper, we propose \method, a method for multi-objective drug molecule discovery. 
\method is based on the idea of generating the chemical candidates by iteratively editing fragments of molecular graphs. 
To search for high-quality candidates, it employs Markov chain Monte Carlo sampling (MCMC) on molecules with an annealing scheme and an adaptive proposal. 
To further improve sample efficiency, \method uses a graph neural network (GNN) to represent and select candidate edits, where the GNN is trained on-the-fly with samples from MCMC. 
Experiments show that \method achieves state-of-the-art performance in various multi-objective settings where molecular bio-activity, drug-likeness, and synthesizability are considered.
Remarkably, in the most challenging setting where all four objectives are simultaneously optimized, our approach outperforms previous methods significantly in comprehensive evaluations. The code is available at \url{https://github.com/yutxie/mars.}

\end{abstract}

\section{Introduction}
\label{sec:intro}
Drug discovery aims to find chemical compounds with desired target properties, such as high drug-likeness~\cite[QED]{bickerton2012quantifying}.
The problem is also referred to as molecular design, molecular generation, or molecular search. 
The space of drug-like chemicals is enormous, approximate $10^{33}$ for realistic drugs that could ever be synthesized~\citep{polishchuk2013estimation}.
Therefore it is very challenging to search for high-quality molecules from such a vast space --- enumeration would take almost forever. 
For a particular disease, finding the right candidates targeting specific proteins further complicates the problem. 

Instead of enumerating or searching from the immense chemical space, recent work utilizes deep generative models to generate candidate molecules directly~\citep{schwalbe-koda2020generative}. However, most prior work focuses on generating molecules concerning a single property such as drug-likeness (QED) or octanol-water partition coefficient (logP)~\citep{jin2018junction,you2018graph,popova2019molecularrnn,shi2020graphaf, zang2020moflow}.
While in practical settings, typical drug discovery requires consideration of multiple properties jointly~\citep{nicolaou2012multi}.
For example, to find drug-like molecules that are easy to synthesize and exhibit high biological activity against the target protein. 
Naturally, multi-objective molecule design is much more challenging than the single-objective scenario~\citep{jin2020composing}. 

This paper studies the problem of multi-objective molecule design for drug discovery.
An ideal solution should be efficient and meet the following criteria. 
\begin{inparaenum}[\it C1:]
    \item It should satisfy multiple properties with high scores; \label{criterion-multi-obj}
    \item It should produce novel and diverse molecules; \label{criterion-div-nov}
    \item Its generation process does not rely on either expert annotated or wet experimental data collected from a biochemistry lab (since it requires tremendous effort and hard to obtain). \label{criterion-data} 
\end{inparaenum}
Existing molecule generation approaches are mainly designed for the single objective setting, and they could not meet all criteria in the setting of multiple objectives. 
These methods belong to four categories: a) generating candidates from a learned continuous latent space~\citep{gomez-bombarelli2018automatic, jin2018junction}, b) through reinforcement learning~\citep{you2018graph},  c) using an encoder-decoder translation approach~\citep{jin2019learning}, or d) optimizing molecular properties through genetic algorithms~\citep{nigam2020augmenting}. 
Current state-of-the-art multi-objective molecular generation is a rationale-based method~\citep{jin2020composing}. In this approach, the authors propose to build molecules by composing multiple extracted rationales, and the model can generate compounds that are simultaneously active to multiple biological targets.
However, such an approach will result in quite complex molecules when we have many objectives. This is because different objectives correspond to different rationales, and including all these rationales could lead to large molecules, which may be less drug-like and hard to be synthesized practically.

In this paper, we propose MArkov moleculaR Sampling (\method), a simple yet flexible method for drug discovery. 
The basic idea is to start from a seed molecule and keep generating candidate molecules by modifying fragments of molecular graphs from previous steps.  
It meets all the criteria C\ref{criterion-multi-obj}-\ref{criterion-data}.
In \method, the molecular design is formulated as an iterative editing procedure with its total objective consisting of multiple property scores (C\ref{criterion-multi-obj}). 
\method employs the annealed Markov chain Monte Carlo sampling method to search for optimal chemical compounds, which allows for the exploration of chemicals with novel and different fragments (C\ref{criterion-div-nov}).
The proposal to modify molecular fragments is represented using graph neural networks (GNNs), whose parameters are adaptively learned. 
We used message passing neural networks (MPNNs) in practice~\citep{gilmer2017neural}, but other GNNs can fit the framework as well.
Furthermore, \method utilizes the sample paths generated on-the-fly to train the proposal network adaptively. Therefore, it does not rely on external annotated data (C\ref{criterion-data}). 
With such an adaptive learnable proposal, it keeps improving the generation quality throughout the process.

We evaluate \method and four other baselines, one latest method for each of the four method categories. 
The benchmark includes a variety of multi-objective generation settings. 
Experiments show that our proposed \method achieves state-of-the-art performance on five out of six tasks in terms of a comprehensive evaluation consisting of the success rate, novelty, and diversity of the generated molecules. 
Notably, in the most challenging setting where four objectives -- bio-activities to two different targets, drug-likeness, and synthesizability -- are simultaneously considered, our method achieves the state-of-the-art result and outperforms existing methods by 77\% in the comprehensive evaluation. 

Our contributions are as follows:
\vspace{-0.6em}
\begin{itemize*}
    \item We present \method, a generic formulation of molecular design using Markov sampling, which can easily accommodate multiple objectives. 
    \item We develop an adaptive fragment-editing proposal based on GNN that is learnable on the fly with only samples self-generated and efficient in exploring the chemical space. 
    \item Experiments verifies our proposed \method framework can find novel and diverse bioactive molecules that are both drug-like and highly synthesizable.
\end{itemize*}

\section{Related Work}
\label{sec:related}
Recent years have witnessed the success of applying deep generative models and molecular graph representation learning in drug discovery~\citep{schwalbe-koda2020generative, guo2020a}. 
Existing approaches for molecular property optimization can be grouped into four categories, including generation with
\begin{inparaenum}[\it a)]
\item Bayesian inference,
\item reinforcement learning, 
\item encoder-decoder translation models, and
\item evolutionary and genetic algorithms. 
\end{inparaenum}
The first category is learning continuous latent spaces for molecular sequences or graphs and generating from such spaces using Bayesian optimization (BO) ~\citep{gomez-bombarelli2018automatic, jin2018junction, winter2019efficient}.
These methods rely heavily on the quality of latent representations, which imposes huge challenges to the encoders when there are multiple properties to consider. 

Unlike the first class, other work uses reinforcement learning (RL) to optimize desired objectives directly in the explicit chemical space~\citep{cao2018molgan,popova2018deep,you2018graph,popova2019molecularrnn,shi2020graphaf}.
However, the models are usually hard to train due to the high variance of RL.

The third category directly trains a translation model that maps from an input molecule to a high-quality output molecule~\citep{jin2019learning, jin2019hierarchical}. 
Although simple, such methods require many high-quality labeled data, making them impractical in scenarios where the data is limited.

The last category of methods are evolutionary algorithms (EAs) and genetic algorithms (GAs) to explore large chemical space with certain property~\citep{nicolaou2012multi, devi2015evolutionary, jensen2019a, ahn2020guiding}. 
In \citet{nigam2020augmenting}, the authors propose to augment GA by adding an adversarial loss into the fitness evaluation to increase the diversity, and the augmented GA outperforms all other generative models in optimizing logP.
Though flexible and straightforward, to make the search process efficient enough, most GA and EA methods require domain experts to design molecular mutation and crossover rules, which could be non-trivial to obtain.

Besides single property optimization, there is recent work to address the multi-objective molecule generation problem.
For example, \citet{li2018multi} proposes to use a conditional generative model to incorporate several objectives flexibly, 
while \citet{lim2020scaffold} leverages molecular scaffolds to control the properties of generated molecules better.
Among them, the current state-of-the-art approach is a rationale-based method proposed by \citet{jin2020composing}. In this method, the authors propose to build molecules by assembling extracted rationales.
Despite its great success in generating compounds simultaneously active to multiple biological targets, the combination of rationales might hinder the synthesizability and drug-likeness of produced molecules, as they tend to be large as the number of objectives grows.
In contrast, our \method framework turns the generation problem into a sampling procedure, which serves as an alternative way compared with deep generative models, and can efficiently discover bio-active molecules that are both drug-like and highly synthesizable.

Remotely related is recent work to generate molecules through sampling.  \citet{seff2019discrete} defines a Gibbs sampling procedure, in which the Markov chain alternates between randomly corrupting the molecules and recovering the corrupted ones with a learned reconstruction model. However, this method mainly focuses on generating molecules that follow the observed data distribution and cannot be directly tailored for property optimization. 
Different from this work, \method is built upon the general MCMC sampling framework, which allows further enhancement with adaptive proposal learning to edit molecular graphs efficiently. 
Actually, generating instances from a discrete space with MCMC sampling methods is previously employed in various other applications, e.g., generating natural language sentences under various constraints~\citep{miao2019cgmh,zhang2019generating,liu-etal-2020-unsupervised,zhang2020language}.

\section{Proposed \method Approach}
\label{sec:method}
In this section, we present the MArkov moleculaR Sampling method (\method) for multi-objective molecular design. 
We define a Markov chain over the explicit molecular graph space and design a kernel to navigate high probable candidates with acceptance and rejection.

\subsection{Sampling from the Molecular Space}
\label{sec:overview}

Our proposed \method framework aims at sampling molecules with desired properties from the chemical space. 
Specifically, given $K$ properties of interest, the desired molecular distribution can be formulated as a combination of all objectives:
\begin{equation}
    \pi(x) = \underbrace{s_1(x)\circ s_2(x)\circ s_3(x)\circ\cdots\circ s_K(x)}_{\text{desired properties}} 
    \label{eq:target_distribution}
\end{equation}
where $x$ is a molecule in the molecular space $\mathcal{X}$. $\pi(x)$ is an unnormalized distribution over molecules integrating the desired properties. $s_k(x)$ is a scoring function for the $k$-th property and the ``$\circ$'' operator stands for a combination of scores (e.g., summation or multiplication). In practical drug discovery, these terms could be related to the biological activity, drug-likeness, and synthesizability of molecules~\citep{nicolaou2012multi}. 
This framework allows flexible configuration according to various concrete applications.
However, as the number of objectives grows, the joint distribution $\pi(x)$ will become more complex and intractable, making the sampling non-trivial. 

In \method, we propose to sample molecules from the desired distribution Eq.~\ref{eq:target_distribution} using Markov chain Monte Carlo (MCMC) methods~\citep{andrieu2003an}. Given a desired molecular distribution $\pi(x)$ as the unnormalized target distribution, we define a Markov chain on the explicit chemical space $\mathcal{X}$ (i.e., each state of the Markov chain is a particular molecule) and introduce a proposal distribution $q(x'\mid x)$ to perform state transitions.

\begin{figure}[ht]
    \centering
    \includegraphics[width=0.9\textwidth]{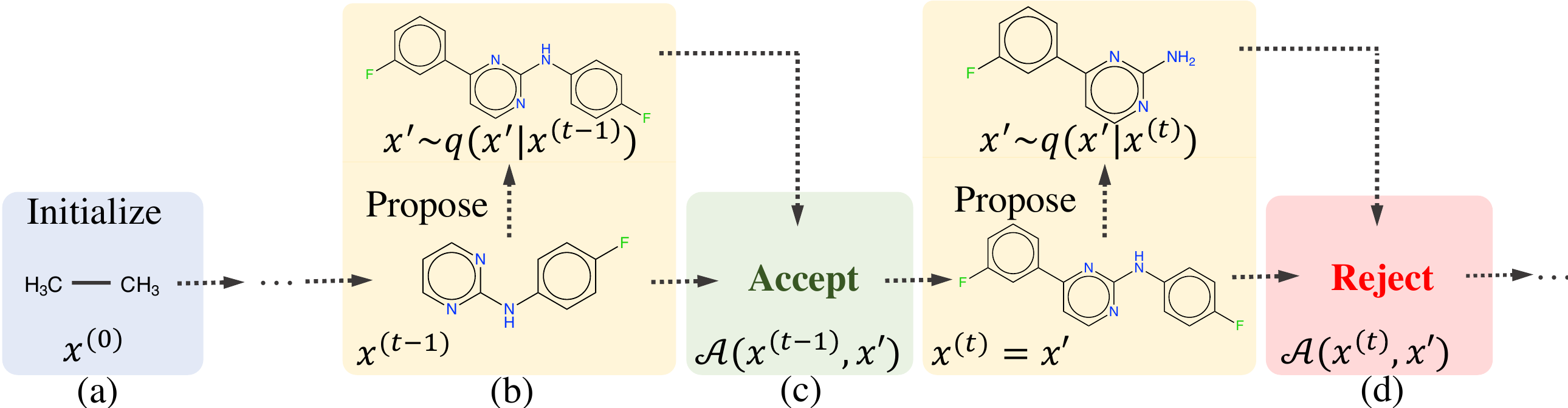}
    % \vspace{+0pt}
    \caption{The framework of \method. During the sampling process: (a) starting from an arbitrary initial molecule $x^{(0)}$ in the molecular space $\mathcal{X}$, (b) sampling a candidate molecule $x'\in\mathcal{X}$ from the proposal distribution $q(x'\mid x^{(t-1)})$ at each step, and (c/d) the candidate $x'$ is either accepted or rejected according to the acceptance rate $\mathcal{A}(x^{(t-1)},x')\in [0,1]$. By repeating this process, we can generate a sequence of molecules $\{x^{(t)}\}_{t=0}^\infty$.
    }
    \label{fig:framework}
    % \vspace{-10pt}
\end{figure}

Specifically, as shown in Figure~\ref{fig:framework}, the sampling procedure of \method starts from an initial molecule $x^{(0)} \in \mathcal{X}$. At each time step $t$, a molecule candidate $x'\in\mathcal{X}$ will be sampled from the proposal distribution $q(x'\mid x^{(t-1)})$, where $x^{(t-1)}$ denotes the molecule at time step $t-1$. Then the proposed candidate $x'$ could be either accepted $x^{(t)}=x'$ or rejected $x^{(t)}=x^{(t-1)}$ according to the acceptance rate $\mathcal{A}(x^{(t-1)},x')\in [0,1]$ controlled by the target distribution $\pi(x)$. By repeating this process, a sequence of molecules $\{x^{(t)}\}_{t=0}^\infty $ can be generated. Such sequence of molecules will converge to the target distribution $\pi(x)$ if the proposal distribution and the acceptance mechanism are configured properly.

The acceptance rate is calculated as follow:
\begin{equation}
    \label{eq:ac}
    \mathcal{A}(x,x') = \min \left \{1,
    \frac{\pi^{\alpha}(x')q(x|x')}{\pi^{\alpha}(x)q(x'|x)} 
    \right \}
\end{equation}
where $\alpha$ is a coefficient that varies in different instantiations of MCMC algorithms. Here to find molecules that globally maximize the target distribution, we employ an annealing scheme~\citep{laarhoven1987simulated} where $\alpha=1/T$ and $T$ is a temperature controlled by a cooling schedule. In addition to this, other instantiations such as Metropolis-Hastings (MH) algorithm~\citep{metropolis1953equation} where $\alpha=1$ are also feasible under our general framework. 

As for the proposal distribution $q(x'\mid x)$, it largely affects the sampling performance and should be designed elaborately. 
In general, it is crucial that the proposal distribution $q(x'\mid x)$ and the target distribution $\pi(x')$ are as close as possible to ensure high sampling efficiency. So we propose using a proposal distribution $q_{\theta}(x'\mid x)$ with learnable parameters to capture the desired molecular properties and develop a strategy to train the proposal throughout the sampling process adaptively. The details will be described in the next section. 

\subsection{Adaptive Molecular Graph Editing Proposal}
\label{sec:proposal}

In this section we will examine in detail our proposed adaptive proposal distribution $q_{\theta}(x'\mid x)$. 
A molecule is represented as a graph whose nodes are heavy atoms and edges are chemical bonds. 
The proposal distribution is defined over molecular graph editing actions. 
We use the message passing neural network (MPNN) to represent the proposal. 
Alternative parameterization schemes such as other graph neural networks are also possible. 
To sample molecules with desired properties effectively and efficiently, we also design a self-training strategy to learn the proposal MPNN during sampling in an adaptive manner.

\begin{figure}[ht]
    \centering
    \includegraphics[width=\textwidth]{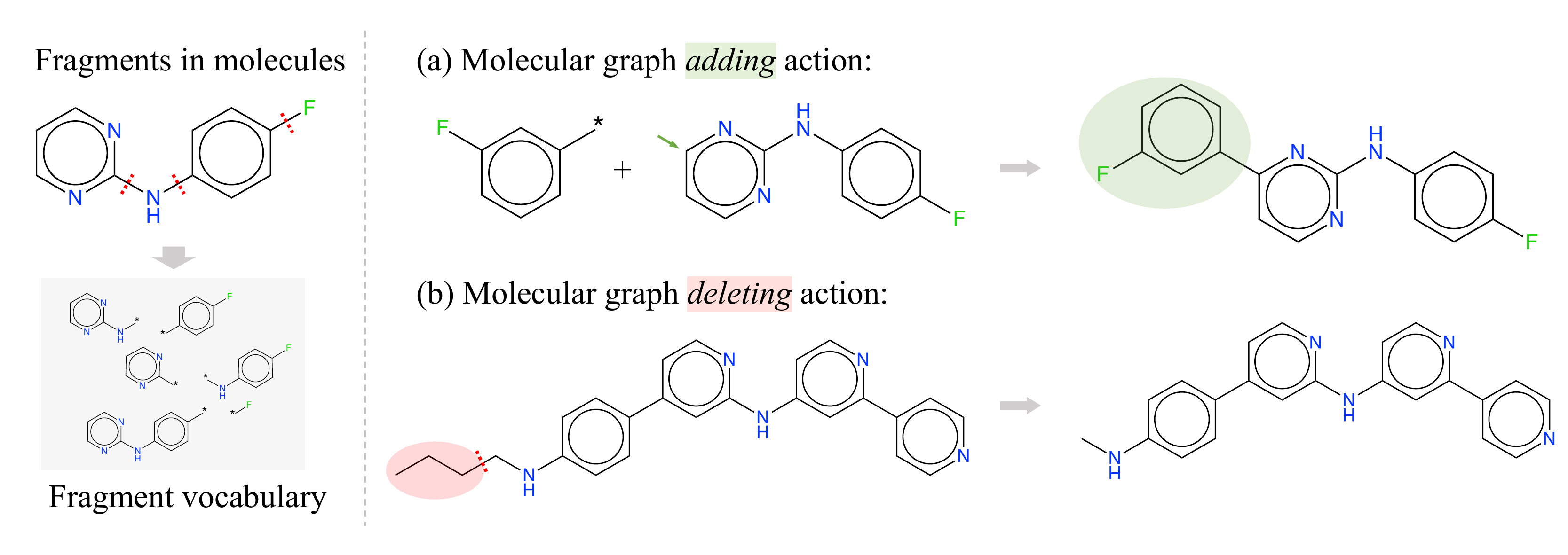}
    \caption{\textbf{Left}: Examples of molecular fragments and a fragment vocabulary. Red dashed lines represents cuttable bonds to extract fragments.
    \textbf{Right}: Examples of molecular graph editing actions. }
    \label{fig:frag-edit}
    % \vspace{-10pt}
\end{figure}

\textbf{Molecular graph editing actions. } 
To transform a molecule $x$ into another molecule $x'$, we consider two sets of graph editing actions, i.e., fragment \emph{adding} and \emph{deleting}. These actions are inspired by fragment-based drug design (FBDD) methodology, whose success in drug discovery has been proved in past decades~\citep{kumar2012fragment}. 
In \method, we define \emph{fragments} as connected components in molecules separated by \emph{single} bonds. To reduce the complexity of editing actions, we only consider fragments with a single attachment position. Moreover, we also define a \emph{fragment vocabulary} that contains finitely many fragments, and only fragments in the vocabulary are allowed to be added onto a molecule. Examples for fragment adding and deleting actions are shown in Figure~\ref{fig:frag-edit}. 

Specifically, given a molecule $x$ with $n$ atoms and $m$ bonds, we choose to add or delete a fragment onto or from this molecule randomly with probability $\frac{1}{2}$ for each set of actions. For the \emph{adding} action, suppose we have a probability distribution over atoms $p_\text{add}(x,u)$ and a probability distribution over fragments in the vocabulary $p_{\text{frag}}(x,u,k)$. Here $u\in [n]$ is an indicator of the atom in $x$ to which the fragment is adding to and $k\in [V]$ is an indicator of fragments in the vocabulary of size $V$. We can compute the proposal distribution as follows: 
\begin{equation}
    \label{eq:proposal_add}
    q(x'|x) = \frac{1}{2}\cdot p_{\text{add}}(x,u)\cdot p_{\text{frag}}(x,u,k)
\end{equation}
where $x'$ is the molecule obtained by adding the $k$-th fragment onto the atom $u$ in molecule $x$. 

As for the \emph{deleting} action, suppose we have a probability distribution over bonds\footnote{Molecular bonds are treated as directional to specify the fragments to drop from molecules.} $p_\text{del}(x,b)$ where $b\in [2m]$ is an indicator of bonds in $x$. We can compute the proposal distribution as follow: 
\begin{equation}
    \label{eq:proposal_del}
    q(x'|x) = \frac{1}{2}\cdot p_{\text{del}}(x,b)
\end{equation}
where $x'$ is the molecule obtained by removing bond $b$ and the attached fragment from molecule $x$. 

\textbf{Parameterizing with MPNNs. } 
To better model the molecular graph editing actions, we propose to use MPNNs to suggest the probability distributions $(p_\text{add}, p_\text{frag}, p_\text{del})=\mathcal{M}_{\theta}(x)$ where $\mathcal{M}_{\theta}$ is a MPNN model specified by parameters $\theta$, which has been proven powerful to predict chemical properties with molecular graphs~\citep{gilmer2017neural}. 
Given a molecule $x$, we compute the probability distributions as follow: 
\begin{align}
    \bm{h}_u^{\text{node}} &= \text{MPNN} (x)_u \in \mathbb{R}^d\\
    \bm{h}_b^{\text{edge}} &= \text{Concat}(\bm{h}_v^{\text{node}}, \bm{h}_w^{\text{node}}) \in\mathbb{R}^{2d} \\
    % \bm{h}^{\text{graph}} &= \text{MaxPooling}(\{\bm{h}_u^{\text{node}}\}) \in\mathbb{R}^d \\
    \label{eq:p_add}
    p_\text{add}(x) &= \text{Softmax}(\{\text{MLP}_{\text{node}}(\bm{h}_u^{\text{node}}))\}_{u=1}^n) \in [0,1]^n \\
    \label{eq:p_frag}
    p_\text{frag}(x,u) &= \text{Softmax}(\text{MLP}_{\text{node}}'(\bm{h}^{\text{node}}_u)) \in [0,1]^{\vert V\vert} \\
    \label{eq:p_del}
    p_\text{del}(x) &= \text{Softmax}(\{\text{MLP}_{\text{edge}}(\bm{h}_b^{\text{edge}}))\}_{b=1}^{2m}) \in [0,1]^{2m}
\end{align}
where $u$ is an atom indicators, $\{\bm{h}_u^{\text{node}}\}_{u=1}^{n}$ are node hidden representations, $v,w$ are atoms connected with bond $b$, $\{\bm{h}_b^{\text{edge}}\}_{b=1}^{2m}$ are edge hidden representations, 
and $\text{MLP}_{\text{node}}, \text{MLP}_{\text{node}}', \text{MLP}_{\text{edge}}$ are multi-layer peceptrons (MLPs), similar to \citet{hu2020strategies}.

\textbf{Adaptive self-training. } 
To capture the desired properties and improve the sampling effectiveness, we can train the editing model to increase the probability of suggesting high-quality candidate molecules. 
Here we propose to train the model on-the-fly during the sampling process in an adaptive manner where the training data is collected from the sampling paths. By doing so, we can bypass the difficulty of lacking training instances that satisfy all property constraints. Mainly, we collect molecule candidates that improve our desired objectives and train the model $\mathcal{M}_{\theta}$ in a maximum likelihood estimation (MLE) manner (i.e., to maximize the probability of producing the collected candidates). The overall  \method is described in Algorithm~\ref{algo:process}.

\begin{algorithm}[t]
\DontPrintSemicolon
    \caption{
    \method
    }
    \label{algo:process} 
    \small
    Set $N$ initial molecules $\{x^{(0)}_i\}_{i=1}^N$ and initialize the molecular graph editing model $\mathcal{M}_{\theta}$\;
    Create an empty editing model training dataset $\mathcal{D}=\{\}$\;
    \For{$t=1,2,\dots$}{
        \For{$i=1,2,\dots, N$}{
            Compute probability distributions $(p_\text{add}, p_\text{frag}, p_\text{del})=\mathcal{M}_{\theta}(x^{(t-1)}_i)$ as Equations~\ref{eq:p_add}-\ref{eq:p_del}\;
            Sample a candidate molecule $x'$ from the proposal distribution $q(x'\mid x^{(t-1)}_i)$ 
            defined with probability distributions $p_\text{add}, p_\text{frag}, p_\text{del}$ 
            as Equations~\ref{eq:proposal_add}-\ref{eq:proposal_del}\;
            \eIf{$u<\mathcal{A}(x^{(t-1)}_i,x')$ where $u\sim\mathcal{U}_{[0,1]}$}{
                Accept the candidate molecule $x^{(t)}_i=x'$\;
            }{
                Refuse the candidate molecule $x^{(t)}_i=x^{(t-1)}_i$\;
            }
            \If{The candidate improves the objectives, i.e. $\pi(x')>\pi(x^{(t-1)}_i)$}{
                Adding the editing record $(x^{(t-1)}_i,x')$ into the dataset $\mathcal{D}$\;
            }
        }
        $\theta^{new} \longleftarrow \arg\max \log M_\theta(\mathcal{D})$ \;
    }
\end{algorithm}

\textbf{Discussion on convergence. }
Compared with standard MCMC algorithms, \method still falls in the Metropolis-Hastings algorithm but with an annealing scheme and an adaptive proposal, which results in inhomogeneous transition kernels.  The convergence of adaptive MCMC is discussed in~\citet{rosenthal2011optimal}. According to the diminishing adaptation condition, we can ensure convergence by making the difference of proposals in consecutive iterations diminish to zero. \method can satisfy this condition by using an optimizer whose learning rate will shrink to zero eventually (e.g., Adam). 
Annealed MCMC is to find samples maximizing the target probability. 
The convergence of annealed MCMC is discussed in~\citet{andrieu2003an}.

\section{Experiments}
\label{sec:experiment}
\newcommand{\graycell}{\cellcolor{gray!25}}

\subsection{Experiment Setup}

\textbf{Biological objectives.}
Following \citet{jin2020composing}, we consider the following inhibition scores against two Alzheimer-related target proteins as the biological activity objectives. The score is given by a random forest model~\footnote{\url{https://github.com/wengong-jin/multiobj-rationale}} that predicts based on Morgan fingerprint features of a molecule~\citep{rogers2010extended}. 
\begin{itemize*}
    \item GSK3$\beta$: Inhibition against glycogen synthase kinase-3$\beta$. 
    \item JNK3: Inhibition against c-Jun N-terminal kinase-3.
\end{itemize*}

\textbf{Non-biological objectives. }
Following \citet{jin2020composing}, we adopt QED~\citep{bickerton2012quantifying} and synthetic accessibility (SA)~\citep{ertl2009estimation} to quantify the drug-likeness and synthesizability. 
We rescale the SA score (initially between $10$ and $1$) into $[0, 1]$ such that molecules with higher scores are more synthesizable.

\textbf{Multi-objective generation setting.}
To evaluate the effectiveness of the proposed method for multi-objective drug design, we also consider the following more challenging objective combinations:
\begin{itemize*}
    \item GSK3$\beta$+JNK3: Jointly inhibiting GSK3$\beta$ and JNK3. The combination may provide potential benefits for the treatment of Alzheimer's disease reported by~\citet{hu2009gsk3,martin2013tau}.
    \item GSK3$\beta$/JNK3+QED+SA: Inhibiting GSK3$\beta$ or JNK3 while being drug-like and synthetically accessible, which are quantified by QED and SA, respectively.
    \item GSK3$\beta$+JNK3+QED+SA: Jointly inhibiting GSK3$\beta$ and JNK3 while being drug-like and synthetically accessible, which are quantified by QED and SA, respectively.
\end{itemize*}

\textbf{Baselines.} We compare \method with the following methods -- the latest ones from four categories mentioned in the related work (Sec.~\ref{sec:related}). 
\textbf{GCPN} \citep{you2018graph} leverages RL to generate molecules atom by atom, and the adversarial loss is incorporated in the objective to generate more realistic molecules. \textbf{JT-VAE} \citep{jin2018junction} is a VAE-based approach that firstly generates junction trees and then assembles them into molecules. It performs Bayesian optimization (BO) to guide molecules towards desired properties. 
\textbf{RationaleRL} \citep{jin2020composing} is a state-of-the-art approach for multi-property optimization, which generates molecules from combined rationales.
\textbf{GA+D} \citep{nigam2020augmenting} is a heuristic search method that applies the genetic algorithm (GA) to find molecules with high property scores. An adversarial loss is incorporated in the fitness evaluation to increase the diversity of generated molecules.

\textbf{Evaluation metrics.} Following \citet{jin2020composing}, we generate $N=5000$ molecules for each approach and compare the proposed method with the baselines on the following evaluation metrics: \textbf{Success rate (SR)} is the percentage of generated molecules that are evaluated as positive on all given objectives (QED $\ge 0.6$, SA $\ge 0.67$, the inhibition scores of GSK3$\beta$ and JNK3 $\ge 0.5$); \textbf{Novelty (Nov)} is the percentage of generated molecules with similarity less than 0.4 compared to the nearest neighbor $x_\text{SNN}$ in the training set \citep{olivecrona2017molecular}: $\text{Nov} = \frac{1}{n}\sum_{x\in\mathcal{G}}\boldsymbol{1}[\text{sim}(x,x_\text{SNN}) < 0.4 ]$; \textbf{Diversity (Div)} measures the diversity of generated molecules, which can be calculated based on pairwise Tanimoto similarity over Morgan fingerprints $\text{sim}(x,x')$ as $\text{Div} = \frac{2}{n(n-1)}\sum_{x\ne x'\in\mathcal{G}}1-\text{sim}(x,x')$; \textbf{PM} is the product of the above three metrics, which is a more comprehensive evaluation of the proposed method.
Intuitively, PM presents the percentage of generated molecules that are simultaneously bio-active, novel and diverse, which are essential criteria for molecules to be considered in building a suitable drug candidate library in early-stage drug discovery~\citep{huggins2011rational}.

\textbf{Implementation details. } 
For the fragment vocabulary, we extract the top 1000 frequently appearing fragments that contain no more than $10$ heavy atoms from the ChEMBL database~\citep{gaulton2017the} by enumerating single bonds to break.
As for the sampling process, the unnormalized target distribution is set as $\pi(x)=\sum_k s_k(x)$ where $s_k(x)$ is a scoring function for the above-mentioned properties of interests, the temperature is set as $T=0.95^{\lfloor t/5\rfloor}$ and we sample $N=5000$ molecules at one time.
During sampling, the computation of $q(x\mid x')$ is ignored and we approximate $\mathcal{A}(x,x')$ with $\min\{1,\pi^\alpha(x')/\pi^\alpha(x)\}$ to increase the computation efficiency. This is acceptable because in practice $q(x\mid x')$ and $q(x'\mid x)$ is of order $O(1)$ and $\mathcal{A}(x,x')$ will be gradually bounded by $\pi^\alpha(x')/\pi^\alpha(x)$ as the temperature $T$ decrease to zero. 
The sampling paths are all starting with an identical molecule ``C-C'', which is also adopted by previous graph generation methods for organic molecules~\citep{you2018graph}.
The MPNN model has six layers, and the node embedding size is $d=64$. 
Moreover, for the model training, we use an Adam optimizer~\citep{kingma2015adam} to update the model parameters with an initial learning rate set as $3\times 10^{-4}$, the maximum dataset size is limited as $\vert \mathcal{D}\vert\le 75,000$, and at each step, we update the model for no more than 25 times.

\subsection{Main Results and Analysis}

We perform ten independent runs for \method. The quantitative results are summarized in Table~\ref{tab:single_two} and Table~\ref{tab:three_four}. From these tables, we observe that \method outperforms all the baselines on five out of six tasks in terms of PM. 
Furthermore, on the most challenging multi-objective optimization task, i.e., GSK3$\beta$+JNK3+QED+SA, it significantly surpasses the best baseline with a 77\% improvement for the product of metrics PM. Additional results are shown in Appendix~\ref{appendix:distribution}. 

\begin{table}[htbp]
    \caption{Comparison of different methods on molecular generation with only bio-activity objectives. Results of GA+D are obtained by running its open-source code. Results of other baselines are taken from \citet{jin2020composing}.
    For \method, we report the mean and standard deviation of 10 independent experiments.
    }
    \label{tab:single_two}
    \vspace{-5pt}
    \begin{center}
    \resizebox{\columnwidth}{!}{
    \begin{tabular}{l|rrrr|rrrr|rrrr}
    \toprule
    \multirow{2}{*}{Method} & 
    \multicolumn{4}{c|}{ GSK3$\beta$ } & 
    \multicolumn{4}{c|}{JNK3} &
    \multicolumn{4}{c}{GSK3$\beta$  + JNK3} \\
     & SR & Nov & Div & PM
     & SR & Nov & Div & PM
     & SR & Nov & Div & PM\\
     
    \midrule
    GCPN & 42.4\% & 11.6\% & 0.904 & \graycell 0.04 & 32.3\% & 4.4\% & 0.884 & \graycell 0.01 & 3.5\% & 8.0\% & 0.874 & \graycell 0.00\\
    
    JT-VAE & 32.2\% & 11.8\% & 0.901 & \graycell 0.03 & 23.5\% & 2.9\% & 0.882 & \graycell 0.01 & 3.3\% & 7.9\% & 0.883 & \graycell 0.00\\
    
    RationaleRL & 100.0\% & 53.4\% & 0.888 & \graycell 0.47 & 100.0\% & 46.2\% & 0.862 & \graycell 0.40 & 100.0\% & 97.3\% & 0.824 & \graycell \textbf{0.80}\\
    GA+D & 84.6\% & 100.0\% & 0.714 & \graycell \textbf{0.60} & 52.8 \% & 98.3\% & 0.726 & \graycell 0.38 & 84.7\% & 100.0\% & 0.424 & \graycell 0.36\\
    \midrule
    % \method-old & 97.8\% & 82.1\% & 0.828 & \graycell \textbf{0.66} & 98.1\% & 85.4\% & 0.774 & \graycell \textbf{0.65} & 99.2\% & 89.6\% & 0.734 & \graycell 0.65\\
    \method & 100.0\% & 84.0\% & 0.718 & \graycell \textbf{0.60} & 98.8\% & 88.9\% & 0.748 & \graycell \textbf{0.66}  & 99.5\% & 75.3\% & 0.691 & \graycell 0.52 \\
    % & $\pm$ 0.0\% & $\pm$ 2.3\% & $\pm$ 0.029 & \graycell $\pm$ 0.04 & $\pm$ 0.3\% & $\pm$ 3.8\% & $\pm$ 0.13 & \graycell $\pm$ 0.04 & $\pm$ 0.1\% & $\pm$ 9.4\% & $\pm$ 0.036 & \graycell $\pm$ 0.09 \\
    & & & & \graycell $\pm$ 0.04 & & & & \graycell $\pm$ 0.04 & & & & \graycell $\pm$ 0.08 \\
    \bottomrule
    \end{tabular}}
    \end{center}
    % \vspace{-6pt}
\end{table}

\begin{table}[htb]
    \caption{Comparison of different methods on molecular generation with bio-activity, QED, and SA objectives. Results of all baselines are obtained by running their open-source codes.
    For the results of \method, we report the mean and standard deviation of 10 independent experiments.
    }
    \label{tab:three_four}
    \vspace{-5pt}
    \begin{center}
    \resizebox{\columnwidth}{!}{
    \begin{tabular}{l | rrrr | rrrr | rrrr}
    \toprule
    \multirow{2}{*}{Method} & 
    \multicolumn{4}{c|}{ GSK3$\beta$ + QED + SA } & 
    \multicolumn{4}{c|}{ JNK3 + QED + SA }  &
    \multicolumn{4}{c}{ GSK3$\beta$ + JNK3 + QED + SA } \\
     & SR & Nov & Div & PM
     & SR & Nov & Div & PM 
     & SR & Nov & Div & PM\\
    \midrule
    GCPN & 0.0\% & 0.0\% & 0.000 & \graycell 0.00 & 0.0\% & 0.0\% & 0.000 & \graycell 0.00  & 0.0\% & 0.0\% & 0.000 & \graycell 0.00\\
    
    JT-VAE & 9.6\% & 95.8\% & 0.680 & \graycell 0.06 & 21.8\% & \textbf{100.0\%} & 0.600 & \graycell 0.13 & 5.4\% & \textbf{100.0\%} & 0.277 & \graycell 0.02\\
    
    RationaleRL & 69.9\% & 40.2\% & \textbf{0.893} & \graycell 0.25 & 62.3\% & 37.6\% & \textbf{0.865} & \graycell 0.20 & 75.0\% & 55.5\% & 0.706 & \graycell 0.29\\
    GA+D & 89.1\% & \textbf{100.0\%} & 0.682 & \graycell 0.61 & 85.7\% & {99.8\%} & 0.504 & \graycell 0.43 & 85.7\% & \textbf{100.0\%} & 0.363 & \graycell 0.31\\
    \midrule
    \method & \textbf{99.5\%} & 95.0\% & 0.719 & \graycell \textbf{0.68} & \textbf{91.3\%} & 94.8\% & 0.779 & \graycell \textbf{0.67} & \textbf{92.3\%} & 82.4\% & \textbf{0.719} & \graycell \textbf{0.55}\\
    % & $\pm$ 0.1\% & $\pm$ 5.7\% & $\pm$ 0.052 & \graycell $\pm$ 0.03 & $\pm$ 2.0\% & $\pm$ 1.9\% & $\pm$ 0.09 & \graycell $\pm$ 0.02 & $\pm$ 4.6\% & $\pm$ 5.3\% & $\pm$ 0.011 & \graycell $\pm$ 0.05 \\
    &  &  &  & \graycell $\pm$ 0.03 & & & & \graycell $\pm$ 0.02 & & & & \graycell $\pm$ 0.05 \\
    \bottomrule
    \end{tabular}
    }
    \end{center}
    % \vspace{-6pt}
\end{table}

In comparing all these methods, the GA+D baseline is most similar to our \method in terms of the high novelty and PM score, as both methods focus on molecular space exploration. However, the diversity score of GA+D drops a lot when optimizing multiple properties simultaneously, as GAs are likely to get trapped in regions of local optima~\citep{paszkowicz2009properties}.
RationaleRL is a very strong baseline that performs better than \method in the GSK3$\beta$+JNK3 setting. Nevertheless, when taking the drug-likeness and synthetic accessibility into consideration, their performance falls short of ours and fails to generate novel molecules. 
The performance of GCPN and JT-VAE remains relatively low in most settings, as they are not tailored for multi-objective property optimization.

\begin{figure}[ht]
\centering
\begin{subfigure}{0.32\textwidth}
    \centering
    \includegraphics[width=\textwidth]{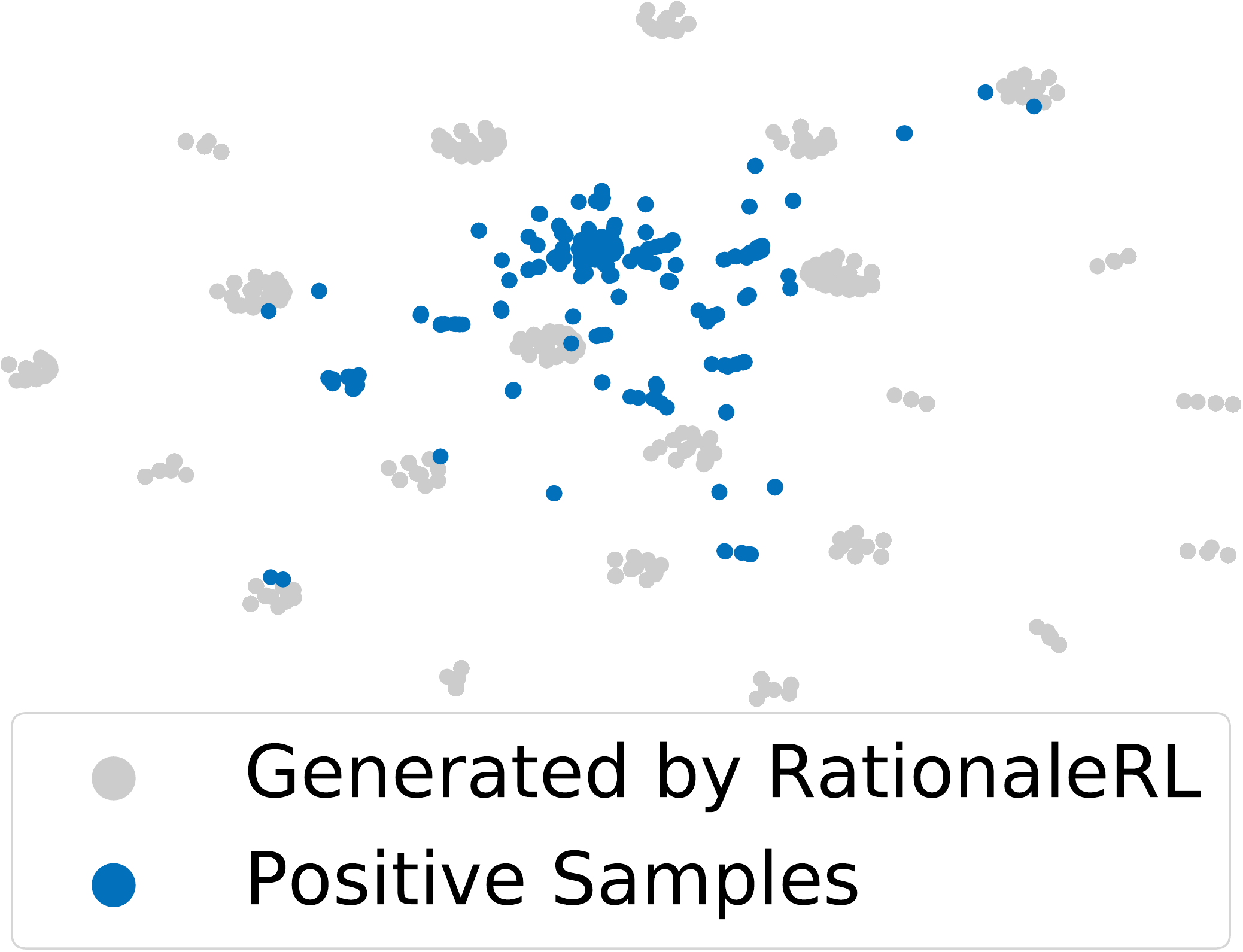}
    \caption{RationaleRL}
\end{subfigure}
\begin{subfigure}{0.32\textwidth}
    \centering
    \includegraphics[width=\textwidth]{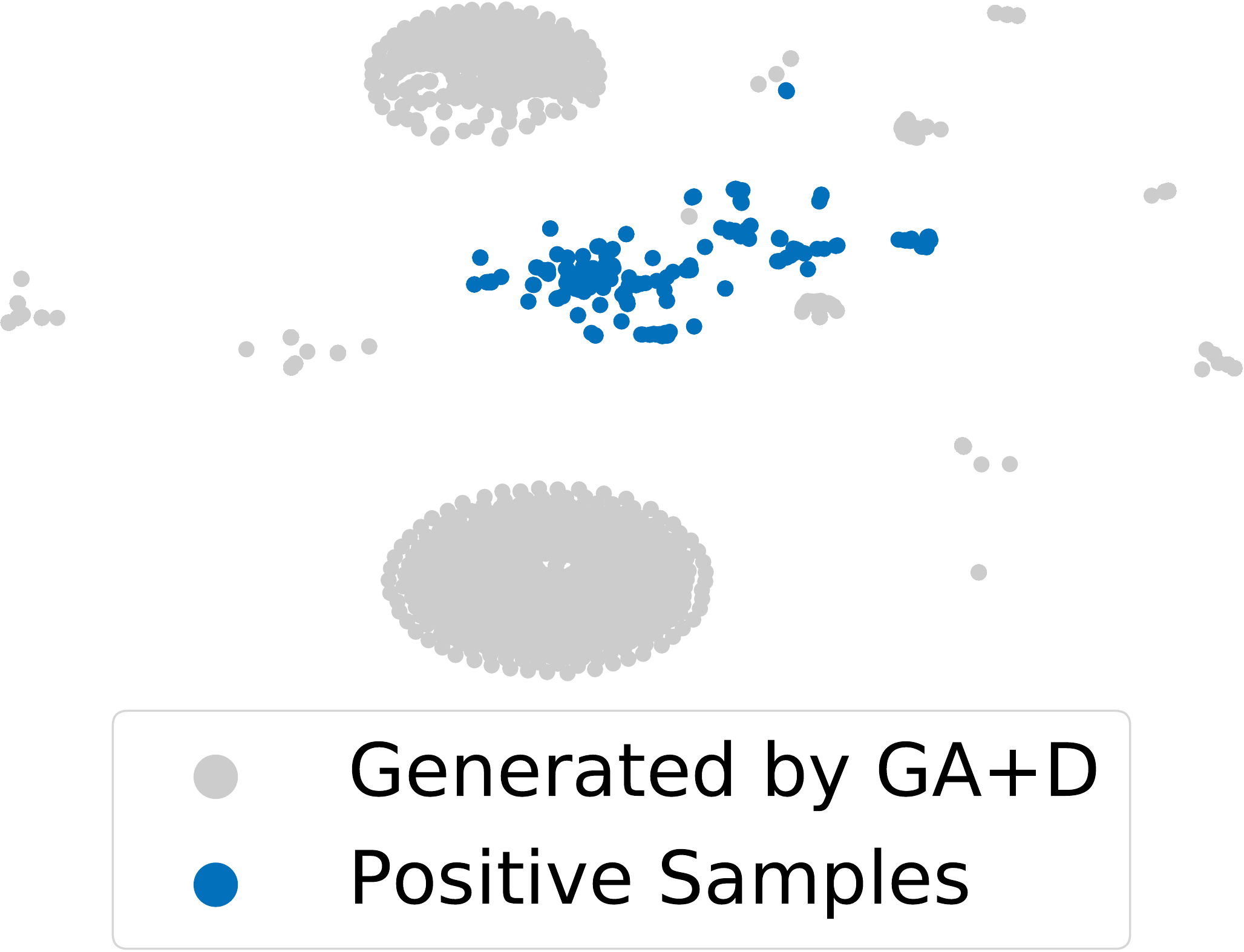}
    \caption{GA+D}
\end{subfigure}
\begin{subfigure}{0.32\textwidth}
    \centering
    \includegraphics[width=\textwidth]{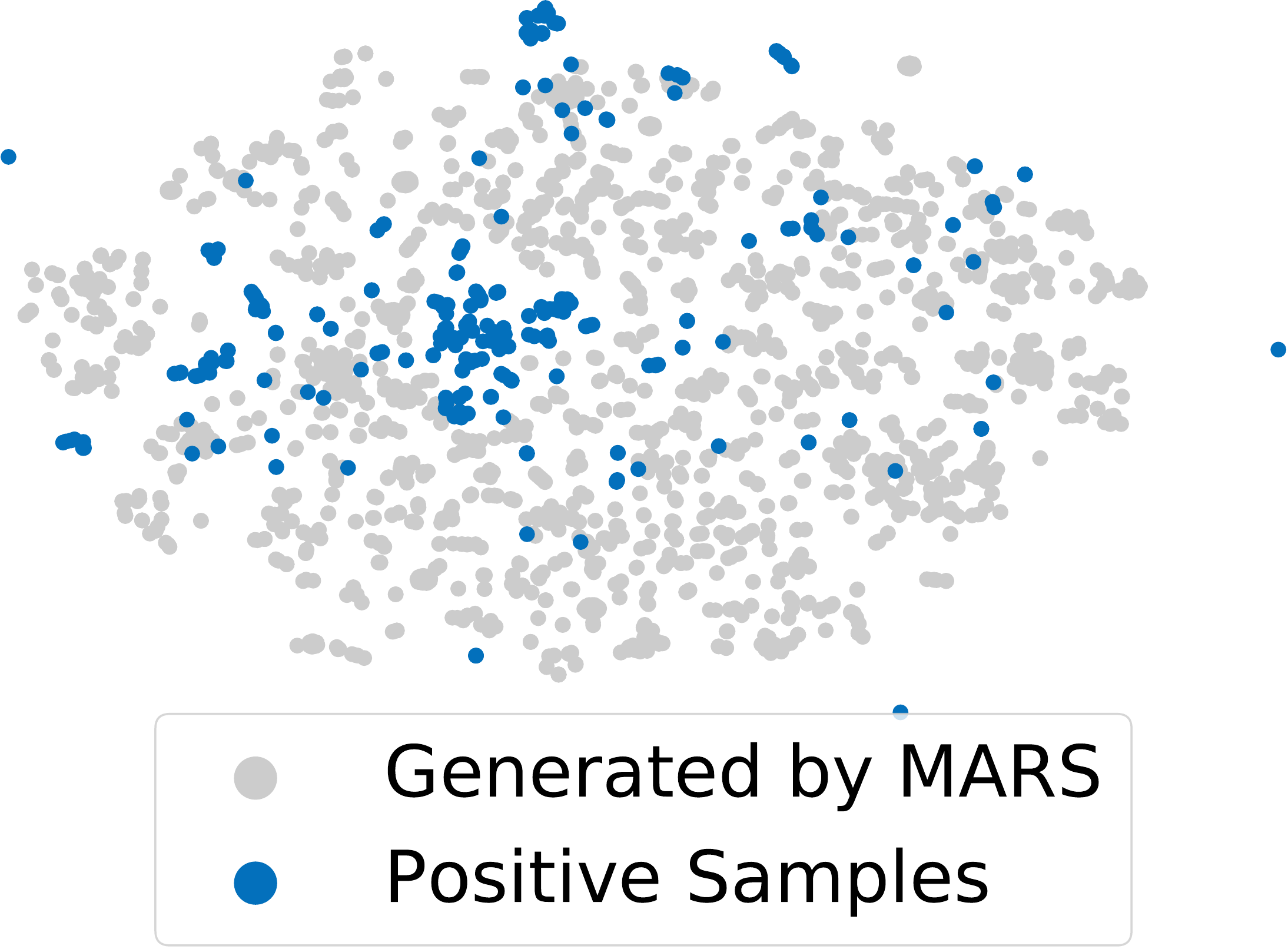}
    \caption{\method}
\end{subfigure}
\caption{t-SNE visualization of generated molecules (gray) and positive molecules in the training set (blue). }
\label{fig:visualization}
% \vspace{-10pt}
\end{figure}

\textbf{Visualization. }
We use t-SNE~\citep{maaten2008visualizing} to visualize the distribution of generated positive molecules with the positive ones in the training set under the GSK3$\beta$+JNK3+QED+SA setting. 
In the visualization, we use the ECFP6 fingerprints as suggested in ~\citet{li2018multi}. 
As shown by Figure~\ref{fig:visualization}, most molecules generated by GA+D fall into two massive clusters, which aligns their low diversity. 
Molecules generated by RationaleRL also tend to be clustered, with each cluster standing for a specific combination of rationales.
By contrast, the molecules generated by \method are evenly distributed in the space with a range of novel regions covered, which justifies our high novelty and diversity scores. We further visualize some molecules generated by \method with high property scores in Figure~\ref{fig:case_good}, indicating its ability to generate highly synthesizable drug-like molecules that jointly inhibit GSK3$\beta$ and JNK3. Additional examples of sampled molecules are shown in Appendix~\ref{appendix:examples}.

\begin{figure}[ht]
\centering
\captionsetup[subfigure]{labelformat=empty}
\begin{subfigure}{0.24\textwidth}
    \centering
    \includegraphics[width=\textwidth]{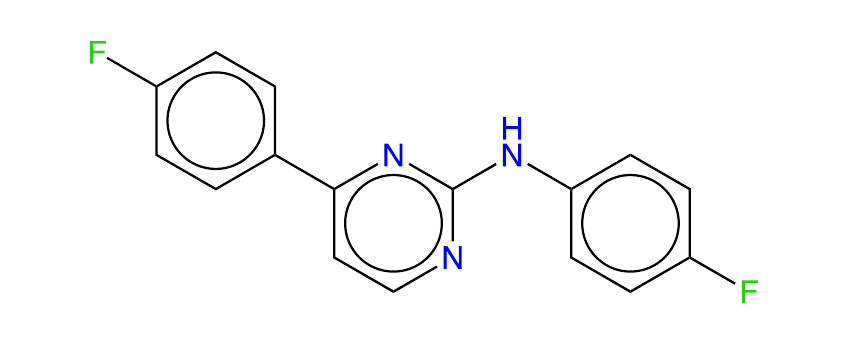}
    \caption{(0.91, 0.85, 0.78, 0.92)}
\end{subfigure}
\begin{subfigure}{0.24\textwidth}
    \centering
    \includegraphics[width=\textwidth]{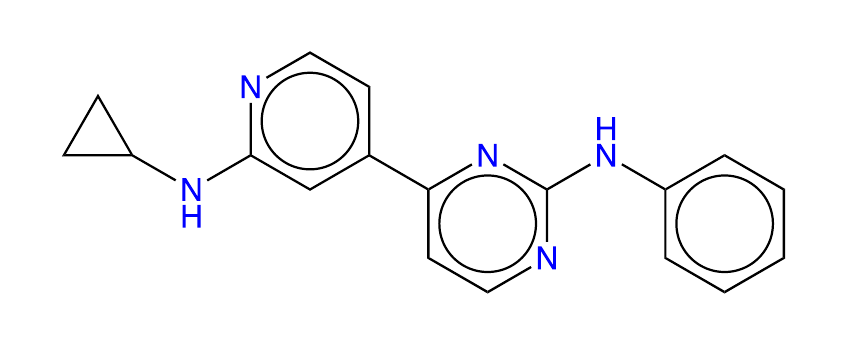}
    \caption{(0.95, 0.76, 0.75, 0.88)}
\end{subfigure}
\begin{subfigure}{0.24\textwidth}
    \centering
    \includegraphics[width=\textwidth]{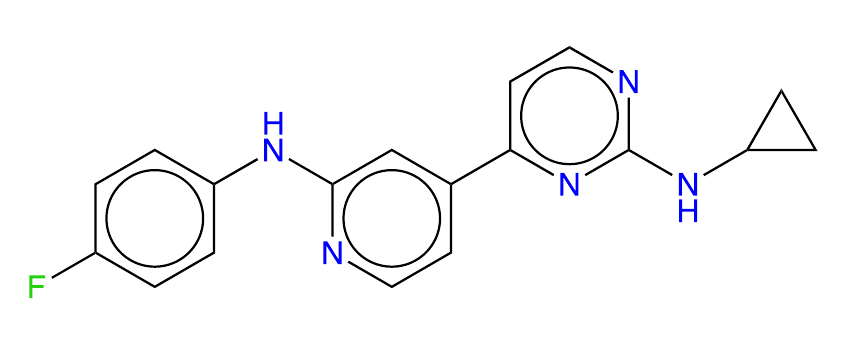}
    \caption{(0.85, 0.87, 0.74, 0.87)}
\end{subfigure}
\begin{subfigure}{0.24\textwidth}
    \centering
    \includegraphics[width=\textwidth]{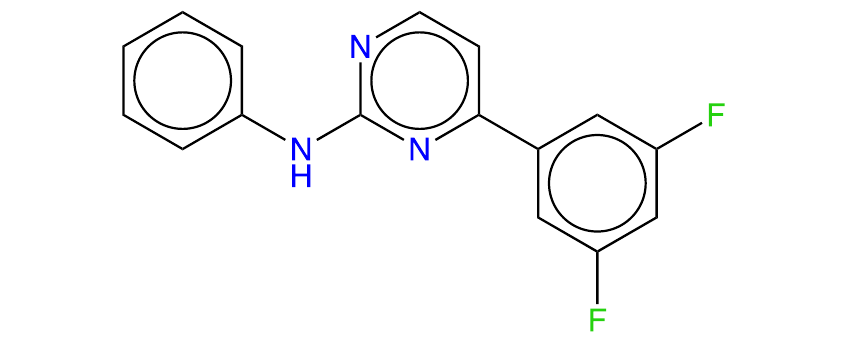}
    \caption{(0.91, 0.71, 0.78, 0.90)}
\end{subfigure}
\caption{Sample molecules generated by \method in the GSK3$\beta$+JNK3+QED+SA setting. The numbers in brackets are GSK3$\beta$, JNK3, QED, and SA scores of each molecule respectively. }
\label{fig:case_good}
% \vspace{-10pt}
\end{figure}

\hide{
\begin{figure}[ht]
    \centering
    \includegraphics[width=\textwidth]{figures/case_good.pdf}
    \caption{Sample molecules generated by \method in the GSK3$\beta$+JNK3+QED+SA setting. The numbers in brackets are GSK3$\beta$, JNK3, QED, and SA scores of each molecule respectively.
    }
    \label{fig:case_good}
    % \vspace{-10pt}
\end{figure}
}

\textbf{Running time. }
The computing server has two CPUs with 64 virtual cores (2.10GHz), 231G memory (about 50G used), and one Tesla V100 GPU with 32G memory.
In the GSK3$\beta$+JNK3+QED+SA setting, MARS takes roughly $T=550$ sampling steps and 12 hours in total to converge (including the time used in proposing and evaluating molecules as well as MPNN model training). 
For other baselines, RationaleRL takes 5.7 hours to fine-tune the model, and GA+D takes 278 steps and 2.2h to achieve its best performance. 
Compared to the conventional drug discovery process, which usually takes months to years, the time we spent on molecular generation models is almost ignorable.

\subsection{Effects of Proposal and Acceptance Strategy}

To justify the contributions of the designed proposal and acceptance strategy, we compare them with some naive ones and summarize the results of different combinations in Table~\ref{tab:ablation}.
For acceptance strategies, 
\textbf{Annealed} stands for annealed MCMC where the acceptance rate is computed as \Eqref{eq:ac} given $\alpha=1/T$, 
\textbf{AlwaysAC} stands for always accepting the candidate, i.e., $\mathcal{A}(x, x')\equiv 1$, and 
\textbf{HillClimb} stands for accepting the candidate only when the overall score is improved, i.e., $\mathcal{A}(x, x')=\operatorname{sign}[s(x')>s(x)]$.
For proposal strategies, 
\textbf{Random} stands for random proposal where we randomly select atoms, bonds, and fragments to edit,
and \textbf{Adaptive} stands for the adaptive fragment-based graph editing model trained during the sampling process as described in Section~\ref{sec:proposal}.

\begin{table}[h]
    \centering
       \caption{
       Results of different acceptance strategies and proposal strategies for molecular sampling. }
       \label{tab:ablation}
       \begin{tabular}{l l | c c c c | c c c c} 
           \toprule
             \multirow{2}{*}{AC Strategy} & 
             \multirow{2}{*}{Proposal} & \multicolumn{4}{c|}{GSK3$\beta$ + JNK3} & \multicolumn{4}{c}{GSK3$\beta$ + JNK3 + QED + SA} \\ 
            &  & SR     & Nov    & Div   & PM   & SR     & Nov    & Div   & PM   \\ 
           \midrule
           Annealed     & Random    & 40.9\% & 94.9\% & 0.828 & \graycell 0.32 & 25.5\% & 80.4\% & 0.793 & \graycell 0.16 \\
           AlwaysAC     & Adaptive  & 49.1\% & 88.4\% & 0.742 & \graycell 0.32 & 10.1\% & 94.6\% & 0.716 & \graycell 0.07 \\ 
           HillClimb    & Adaptive  & 53.7\% & 96.1\% & 0.814 & \graycell 0.42 & 51.4\% & 86.6\% & 0.777 & \graycell 0.35 \\
           Annealed     & Adaptive  & 99.5\% & 75.2\% & 0.688 & \graycell 0.52 & 92.3\% & 82.4\% & 0.719 & \graycell 0.55 \\ 
           \bottomrule
       \end{tabular}
    % \vspace{-7pt}
\end{table}

The results in Table \ref{tab:ablation} indicate that proposals will influence the performance of \method dramatically (the first and the last row), especially when the number of objectives increases. The proposed adaptive proposal outperforms the random proposal and converges 4.6x faster in practice.
By comparing the last three rows, we find the Annealed strategy outperforms the other two strategies by a large margin on both settings, as samples from such strategy are more likely to jump out of local optimums. Another interesting observation is that even with the naive AlwaysAC or heuristic HillClimb strategy, the \method achieves comparable or even better performance than GA+D and RationaleRL in some settings, e.g., HillClimb on GSK3$\beta$+JNK3+QED+SA optimization, which again proves the effectiveness of the proposed proposal.

\section{Conclusion and Future Work}
\label{sec:conclusion}
This paper proposes a simple yet flexible MArkov moleculaR Sampling framework (\method) for multi-objective drug discovery. 
\method includes a trainable proposal to modify chemical graph fragments, which is parameterized by an MPNN.
Our experiments verify that \method outperforms prior approaches on five out of six molecule generation tasks, and it is capable of finding novel and diverse bioactive molecules that are both drug-like and highly synthesizable. Future work can include further study of parameterization and training strategy of the molecular-editing proposal.

% \newpage

\section{Acknowledgement}
We would like to thank Meihua Dang for refactoring much of the \method code. Meihua also performed multiple experiments, which generates the results for the tables. We also thank Jiangjie Chen, Yuxuan Song, Jingjing Xu, Weiying Ma, Hang Li, and anonymous reviewers for their constructive comments and suggestions. 

\bibliography{main}
\bibliographystyle{iclr2021_conference}

% \clearpage
\newpage

\appendix
\appendixpage

\section{Property Scores of Sampled Molecules}
\label{appendix:distribution}
\begin{figure}[ht]
\centering
\begin{subfigure}{0.45\textwidth}
    \centering
    \includegraphics[width=\textwidth]{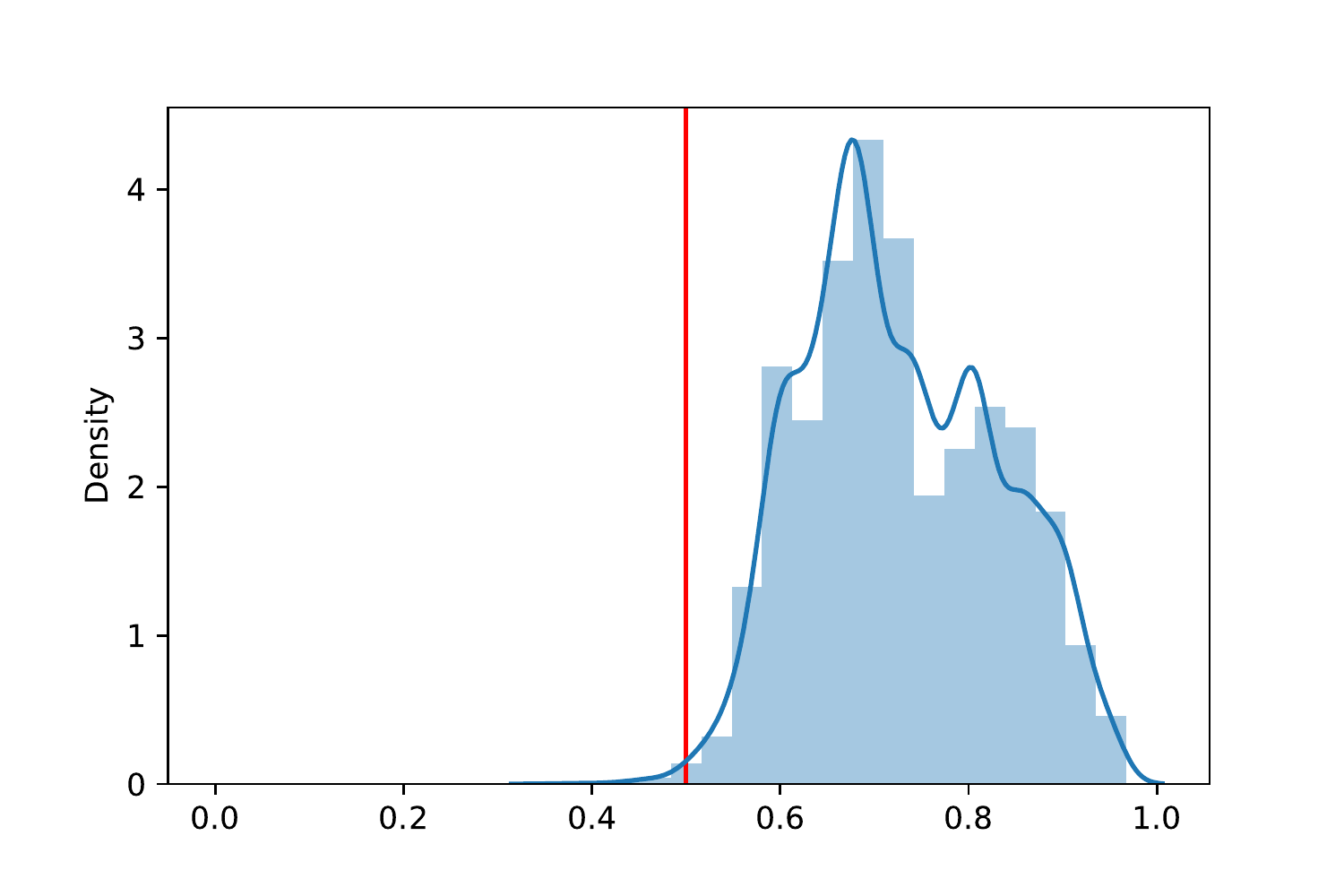}
    \caption{GSK3$\beta$ inhibition score distribution. }
\end{subfigure}
\begin{subfigure}{0.45\textwidth}
    \centering
    \includegraphics[width=\textwidth]{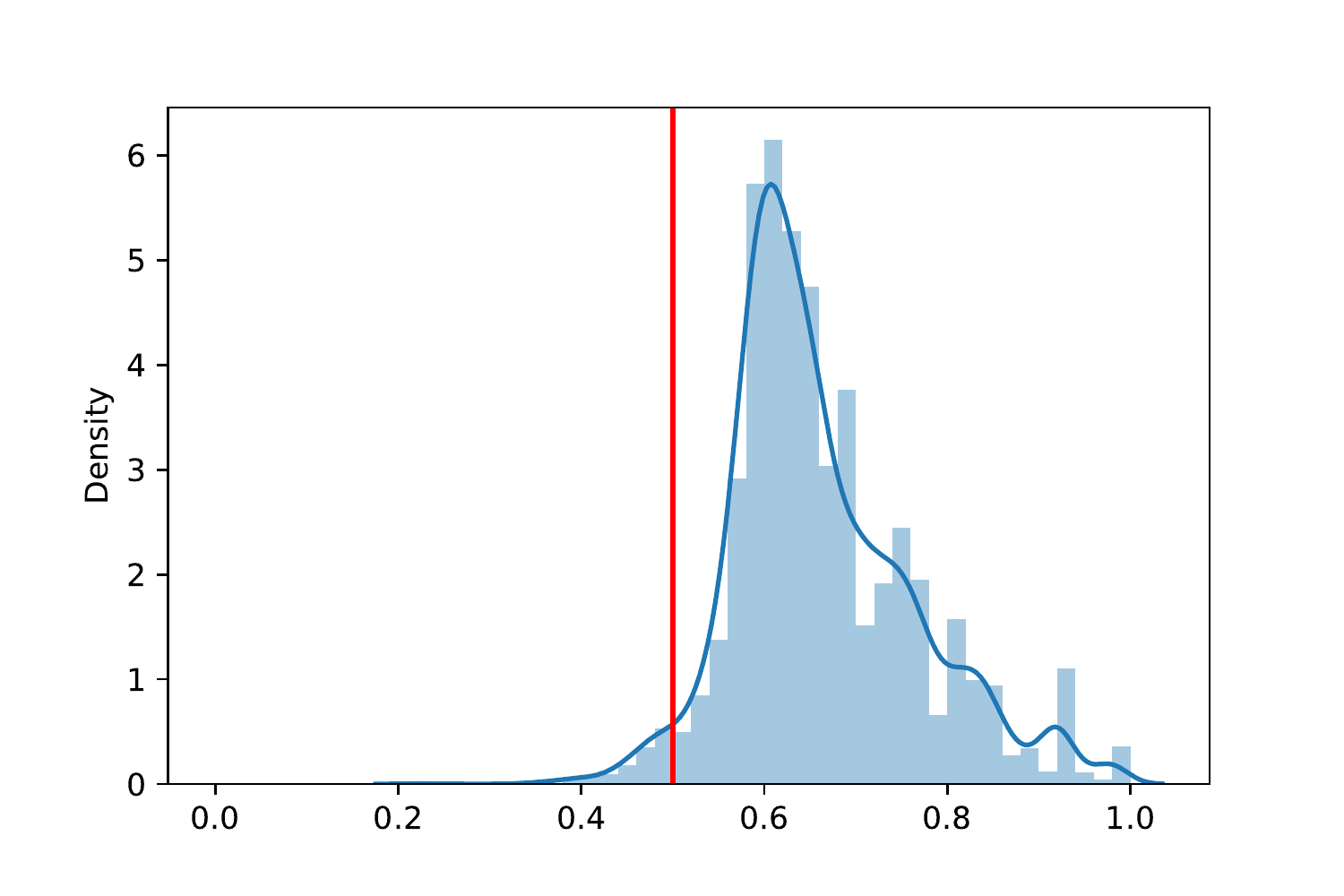}
    \caption{JNK3 inhibition score distribution. }
\end{subfigure}
\bigskip
\begin{subfigure}{0.45\textwidth}
    \centering
    \includegraphics[width=\textwidth]{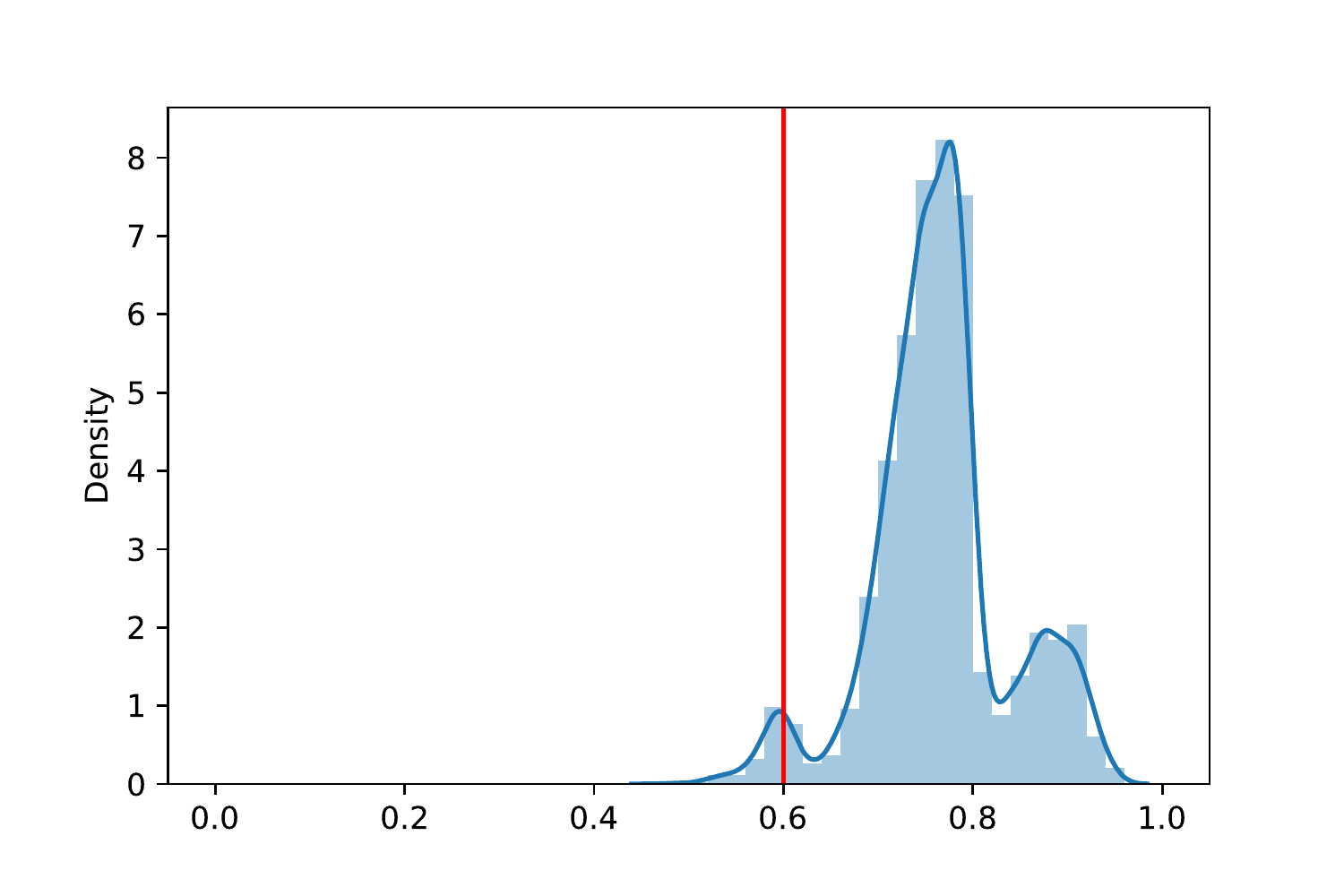}
    \caption{QED score distribution. }
\end{subfigure}
\begin{subfigure}{0.45\textwidth}
    \centering
    \includegraphics[width=\textwidth]{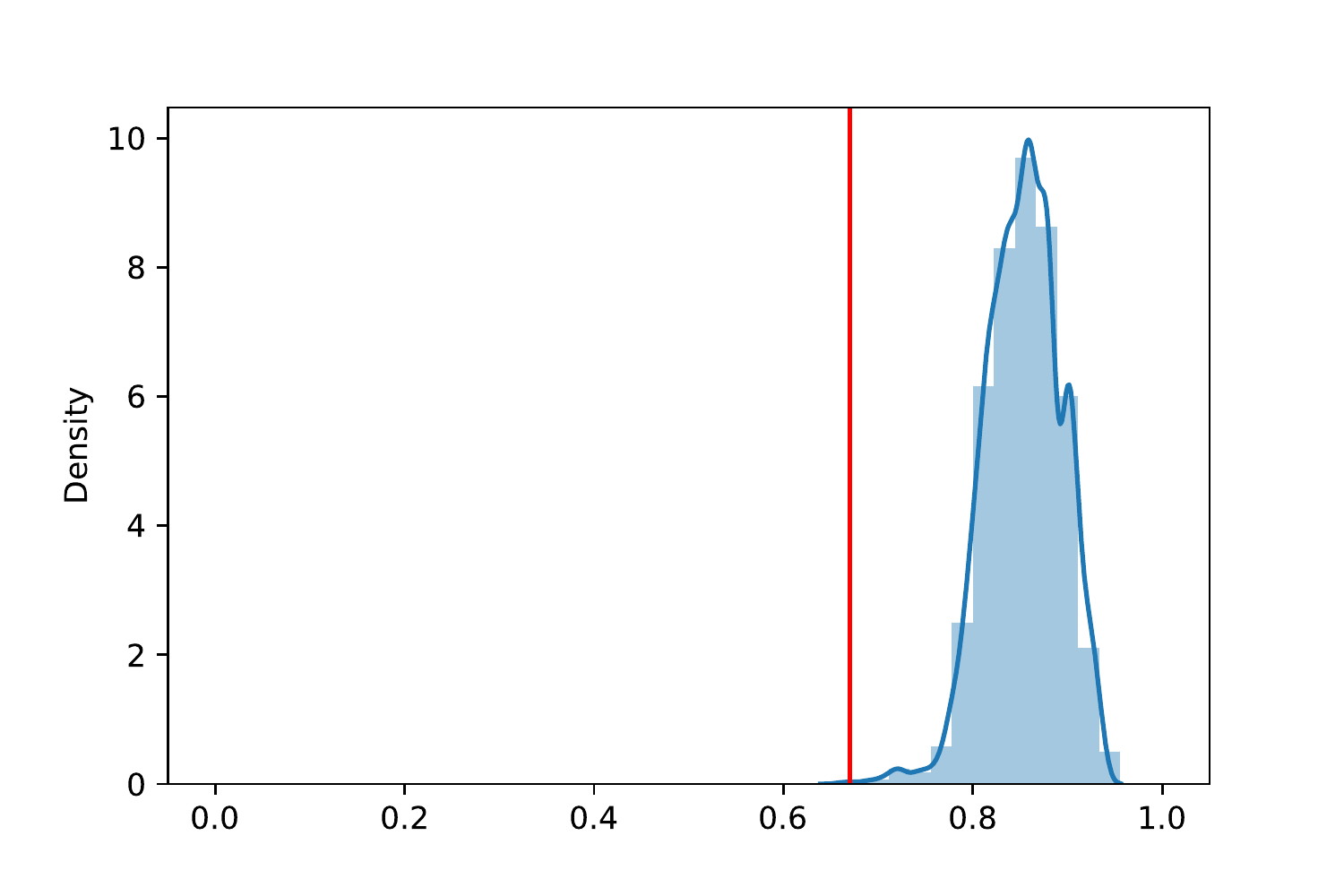}
    \caption{SA score distribution. }
\end{subfigure}
\caption{Property score distributions of sampled $N=5000$ molecules. The red lines are success thresholds. }
\label{fig:dists}
\vspace{-10pt}
\end{figure}

\begin{figure}[ht]
    \centering
    \includegraphics[width=0.6\textwidth]{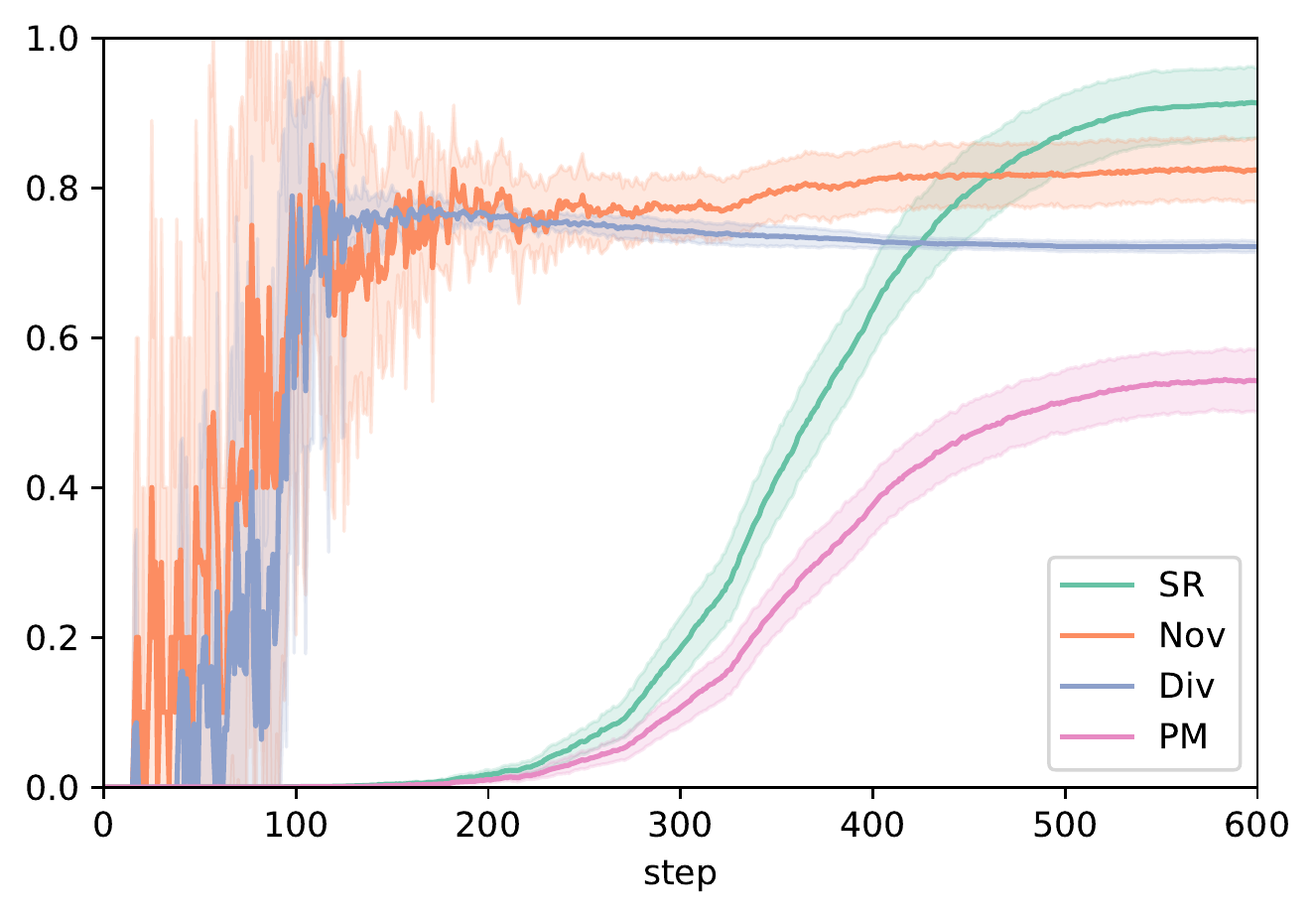}
    \caption{
    \method sampling curves (average of 10 runs) for the GSK3$\beta$+JNK3+QED+SA setting. SR: success rate. Nov: novelty. Div: diversity. PM: product of the three metrics. Shaded area shows the standard deviations over 10 independent runs. 
    }
    \label{fig:curves}
\end{figure}

The property score distributions of sampled $N=5000$ molecules of the GSK3$\beta$+JNK3+QED+SA setting are shown in Figure~\ref{fig:dists}.
The average of the metrics over the sampling path is shown in Figure~\ref{fig:curves}.

\section{Single Objective Generation}
\label{appendix:single}
To study whether our proposed method is capable of single-objective molecular generation, we also investigate how \method performs on the drug-likeness (QED) and the penalized octanol-water partition coefficient (penalized logP) optimization. The experiment results are shown in Table~\ref{tab:single_obj}. In the experiments, our approach can obtain the best performance on both QED and logP optimization. And especially, \method outperforms previous methods significantly in the logP generation task. 

\begin{table}[htbp]
    \caption{Comparison of different methods on single-objective molecular generation. Results of other baselines are taken from \citet{shi2020graphaf} and \citet{nigam2020augmenting}.}
    \label{tab:single_obj}
    \vspace{-5pt}
    \begin{center}
    % \resizebox{\columnwidth}{!}{
    \begin{tabular}{l | ccc | ccc}
    \toprule
    \multirow{2}{*}{Method} & 
    \multicolumn{3}{c|}{ QED } & 
    \multicolumn{3}{c}{ Penalized logP } \\
    & 1st & 2nd & 3rd
    & 1st & 2nd & 3rd \\
     
    \midrule
    GCPN~\citep{you2018graph} & \textbf{0.948} & 0.947 & 0.946 & 7.98 & 7.85 & 7.80 \\
    JT-VAE~\citep{jin2018junction} & 0.925 & 0.911 & 0.91 & 5.30 & 4.93 & 4.49 \\
    GraphAF~\citep{shi2020graphaf} & \textbf{0.948} & \textbf{0.948} & 0.947 & 12.23 & 11.29 & 11.05 \\
    \midrule
    GB-GA~\citep{jensen2019a} & / & / & / & 15.76 $\pm$ 5.71 & / & / \\
    GA+D~\citep{nigam2020augmenting} & / & / & / & 20.72 $\pm$ 3.14 & / & /\\
    \midrule
    \method & \textbf{0.948} & \textbf{0.948} & \textbf{0.948} & 44.99 & 44.32 & 43.81 \\
    \bottomrule
    \end{tabular}
    % }
    \end{center}
    \vspace{-6pt}
\end{table}

Moreover, from the results, we also can see how these two previously widely used metrics~\citep{jin2018junction, you2018graph, popova2019molecularrnn, shi2020graphaf, nigam2020augmenting} are questionable for both scientific study and practical use. Most of the generative methods (i.e., GCPN, JT-VAE, and GraphAF) can produce molecules with the highest possible QED score of 0.948, making the top QED score metric hard to distinguish different methods. As for logP optimization, heuristic search-based (i.e., GB-GA and GA+D) and sampling-based methods (i.e., \method) can all easily beat generative models. This is because penalized logP score will prefer larger molecules that generative models can hardly produce. However, such large molecules are unrealistic for practical drug discovery, making the top penalized logP score metric problematic.

\section{Examples of Sampled Molecules}
\label{appendix:examples}
We also provide some examples of sampled molecules from the GSK3$\beta$+JNK3+QED+SA setting. The numbers under molecule graphs are GSK3$\beta$, JNK3, QED, and SA scores, respectively.

\begin{figure}[ht]
    \centering
    \includegraphics[width=\textwidth]{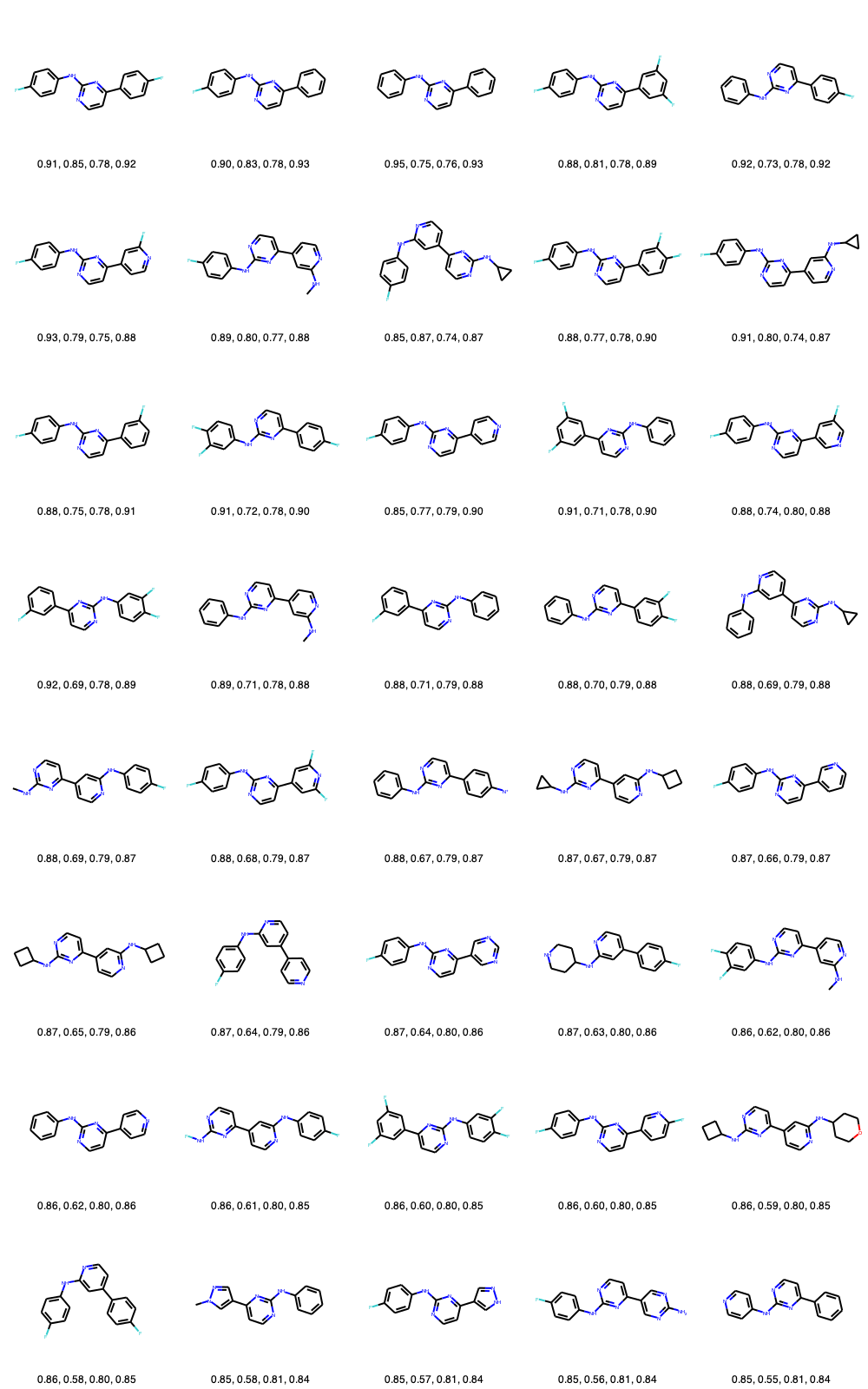}
    \caption{
    40 sampled molecules with highest average property scores. 
    }
    \vspace{-10pt}
\end{figure}

\begin{figure}[ht]
    \centering
    \includegraphics[width=\textwidth]{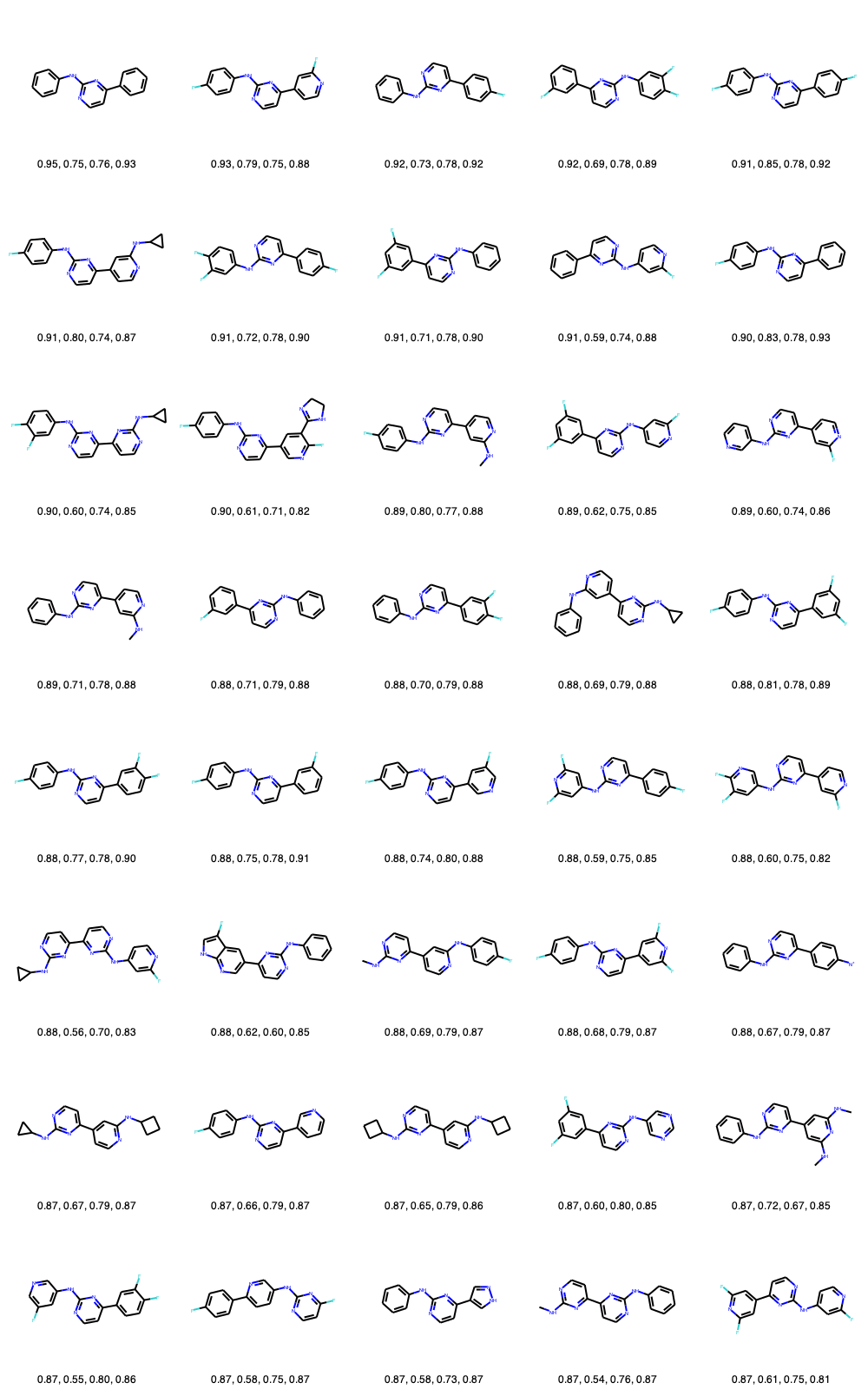}
    \caption{
    40 sampled molecules with highest GSK3$\beta$ scores. 
    }
    \vspace{-10pt}
\end{figure}

\begin{figure}[ht]
    \centering
    \includegraphics[width=\textwidth]{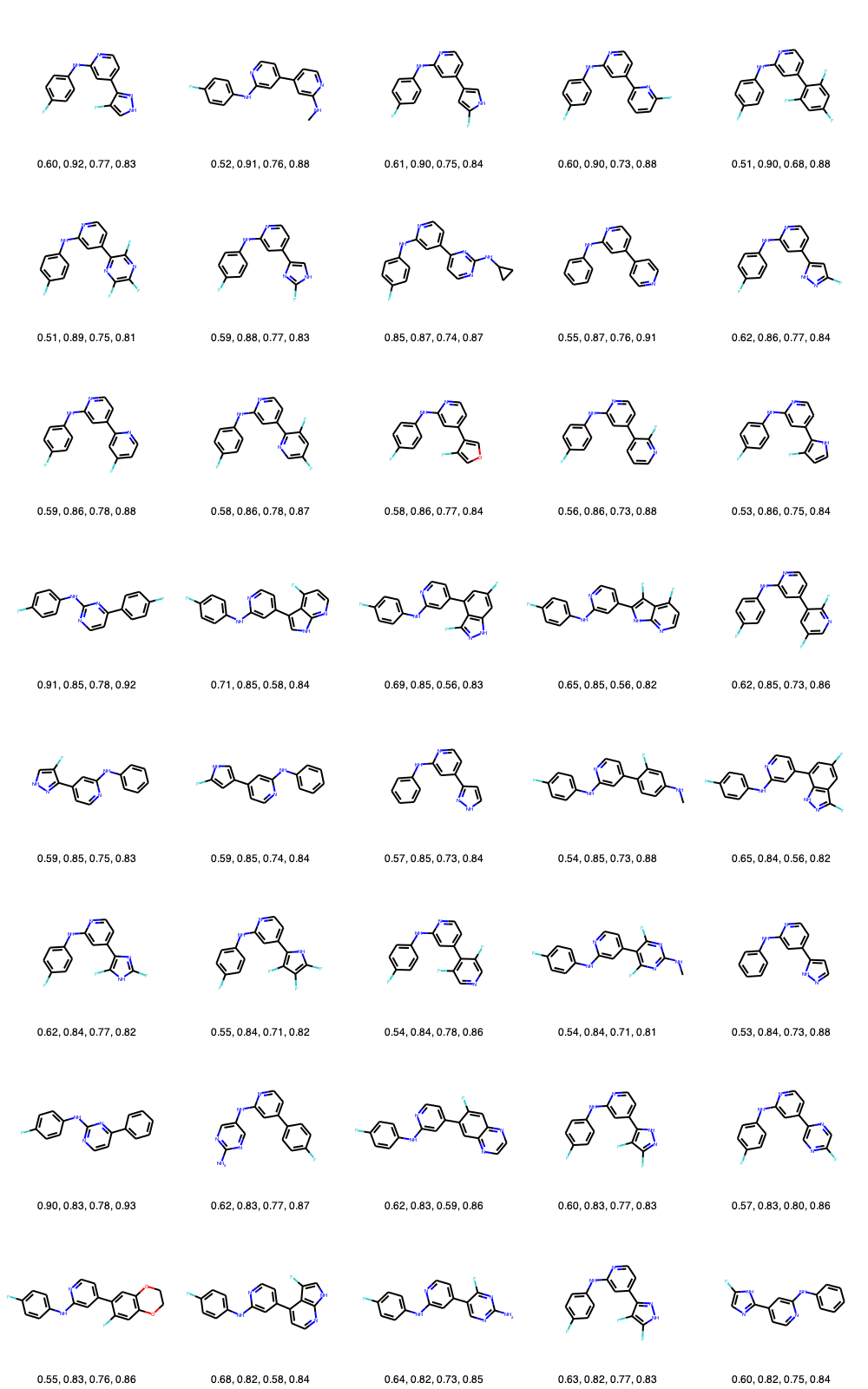}
    \caption{
    40 sampled molecules with highest JNK3 scores. 
    }
    \vspace{-10pt}
\end{figure}

\begin{figure}[ht]
    \centering
    \includegraphics[width=\textwidth]{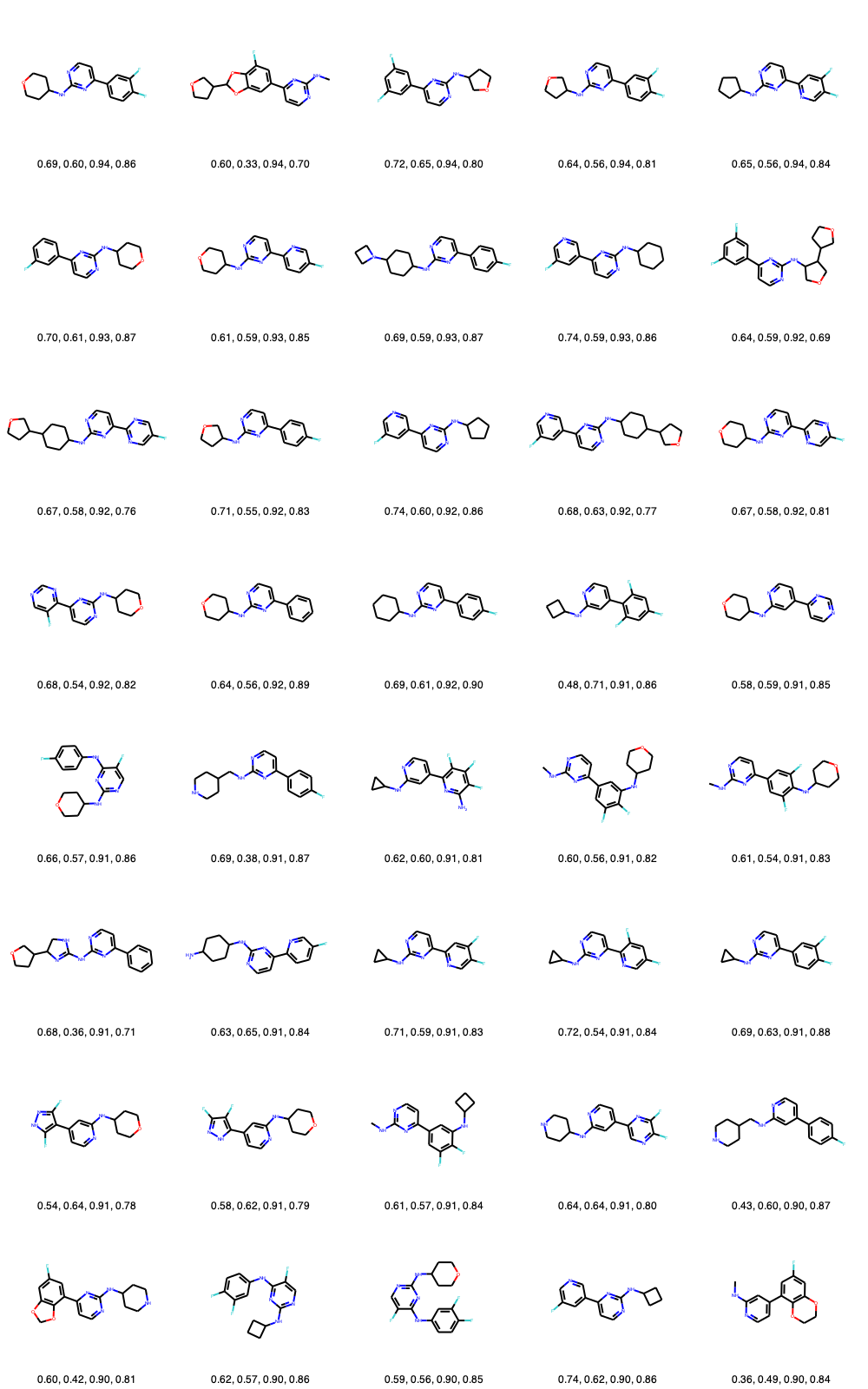}
    \caption{
    40 sampled molecules with highest QED scores. 
    }
    \vspace{-10pt}
\end{figure}

\begin{figure}[ht]
    \centering
    \includegraphics[width=\textwidth]{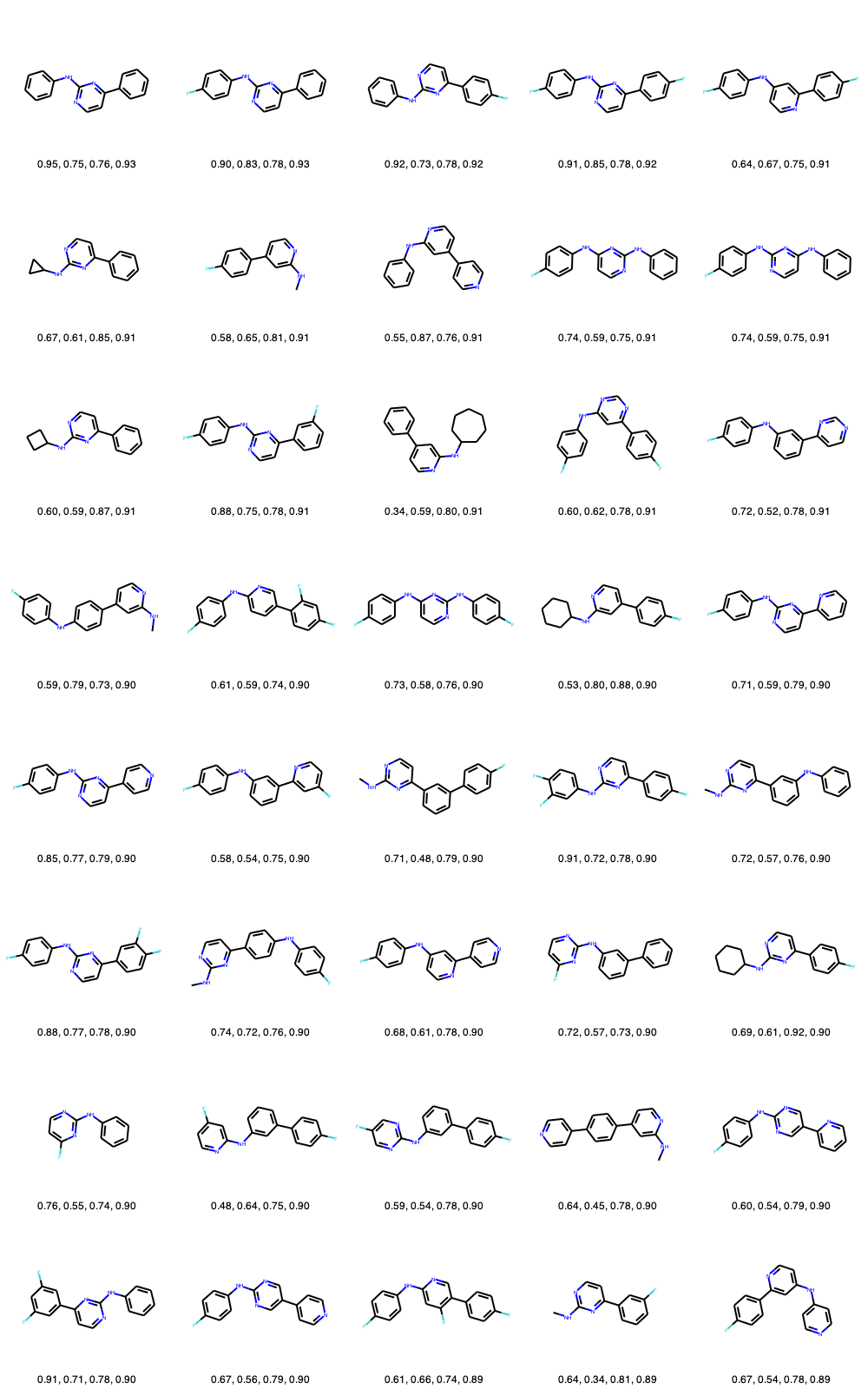}
    \caption{
    40 sampled molecules with highest SA scores. 
    }
    \vspace{-10pt}
\end{figure}

\end{document}